\documentclass[%
 reprint,
 amsmath,amssymb,
 aps,
showkeys
]{revtex4-2}

\usepackage{graphicx}
\usepackage{dcolumn}
\usepackage{multirow}
\usepackage{booktabs}
\usepackage{bm}

\begin{document}

\preprint{APS/123-QED}

\title{Muon Detection and Direction Reconstruction with the Upgraded 1-ton Water Cherenkov Prototype Detector at CJPL-I} 

    \newcommand{\TUDEP}{\affiliation{Department of Engineering Physics \& Center for High Energy Physics, Tsinghua University, Beijing 100084, China}}
    \newcommand{\TUPRI}{\affiliation{Key Laboratory of Particle \& Radiation Imaging (Tsinghua University), Ministry of Education, China}}
    \newcommand{\HNUSPE}{\affiliation{School of Physics \& Electronics, Hunan University, Changsha 410082, China}}
    \newcommand{\HNUPKL}{\affiliation{Hunan Provincial Key Laboratory of High-Energy Scale Physics and Applications,  Changsha 410082, China}}
    \newcommand{\NKU}{\affiliation{School of Physics, Nankai University, Tianjin 300071, China}}
    \newcommand{\UCASP}{\affiliation{School of Physical Sciences, University of Chinese Academy of Sciences, Beijing 100049, China}}
    \newcommand{\SYSUP}{\affiliation{School of Physics, Sun Yat-Sen University, Guangzhou 510275, China}}
    \newcommand{\NUSP}{\affiliation{School of Physics, Nanjing University, Nanjing 210093, China}}
    \newcommand{\SDU}{\affiliation{Institute of Frontier and Interdisciplinary Science, Shandong  University, Qingdao, 266237, China}}
    \newcommand{\JDUFS}{\affiliation{Jinping Deep Underground Frontier Science and Dark Matter Key Laboratory of Sichuan Province, China}}
    \newcommand{\YRHDC}{\affiliation{Yalong River Hydropower Development Company, Ltd., 288 Shuanglin Road, Chengdu 610051, China}}
    \newcommand{\LZU}{\affiliation{School of Nuclear Science and Technology \& MOE Frontiers Science Center for Rare Isotopes, Lanzhou University, Lanzhou 730000, China}}

    \author{Yapeng Wang}\TUDEP\TUPRI
    \author{Yuzi Yang}\email{yangyz18@tsinghua.org.cn}\LZU
    \author{Shaomin Chen}\email{chenshaomin@tsinghua.edu.cn}\TUDEP\TUPRI
    \author{Wei Dou}\TUDEP\TUPRI
    \author{Haoyang Fu}\TUDEP\TUPRI
    \author{Guanghua Gong}\TUDEP\TUPRI
    \author{Lei Guo}\TUDEP\TUPRI
    \author{Ziyi Guo}\TUDEP\TUPRI
    \author{XiangPan Ji}\NKU
    \author{Jianmin Li}\TUDEP\TUPRI
    \author{Jinjing Li}\HNUSPE\TUDEP
    \author{Bo Liang}\TUDEP\TUPRI
    \author{Ye Liang}\TUDEP\TUPRI
    \author{Ling Liu}\TUDEP\TUPRI
    \author{Qian Liu}\UCASP
    \author{Zhiyi Liu}\LZU
    \author{Wentai Luo}\TUDEP\TUPRI
    \author{Ming Qi}\NUSP
    \author{Wenhui Shao}\TUDEP\TUPRI
    \author{Haozhe Sun}\TUDEP\TUPRI
    \author{Jian Tang}\SYSUP
    \author{Yuyi Wang}\TUDEP\TUPRI
    \author{Zhe Wang}\TUDEP\TUPRI
    \author{Changxu Wei}\TUDEP\TUPRI
    \author{Jun Weng}\TUDEP\TUPRI
    \author{Yiyang Wu}\TUDEP\TUPRI
    \author{Benda Xu}\TUDEP\TUPRI
    \author{Chuang Xu}\TUDEP\TUPRI
    \author{Tong Xu}\TUDEP\TUPRI
    \author{Tao Xue}\TUDEP\TUPRI
    \author{Haoyan Yang}\TUDEP\TUPRI
    \author{Aiqiang Zhang}\TUDEP\TUPRI
    \author{Bin Zhang}\TUDEP\TUPRI
    \author{Xinshun Zhang}\TUDEP\TUPRI
    \author{Yang Zhang}\SDU
    \author{Zhicai Zhang}\TUDEP\TUPRI
    \author{Lin Zhao}\TUDEP\TUPRI
    \author{Yangheng Zheng}\UCASP

\collaboration{JNE Collaboration}

\date{\today}

\begin{abstract}
The 1‑ton prototype detector for Jinping neutrino experiment (JNE-1ton) has completed its hardware upgrade and has been successfully operated in both water and liquid‑scintillator modes at CJPL‑I, which is situated under a 2400‑m rock overburden. During the upgrade, the detector’s mechanical support structure and PMT layout were redesigned. The original Hamamatsu PMTs were replaced with 8‑inch MCP‑PMTs from North Night Vision, increasing the PMT count from 30 to 60. The detector was then operated in water mode for 85 days (Water‑I) and 79 days (Water‑II). From the accumulated water-mode data, we derive a muon detection efficiency (Water-II) that is approximately 59\% higher than that of the pre-upgrade liquid scintillator detector. The measured muon flux is $\phi_{\text{I+II}} = (3.55 \pm 0.43_{\mathrm{stat}}\pm 0.28_{\mathrm{syst}}) \times 10^{-10}~\mathrm{cm}^{-2}\mathrm{s}^{-1}$, which is consistent with the previous measurement.Owing to the characteristic angular dependence of Cherenkov radiation and the increased PMT coverage, the uncertainty in muon direction reconstruction is approximately 6$^\circ$, which corresponds to a reduction to 30\% of its previous value. Notably, one rare up‑going muon event was clearly identified. It is excluded from the cosmic-ray muon flux sample and opens a new window for neutrino‑induced event studies in the deepest underground laboratory in China. This work marks the first time that CJPL has employed a cost‑effective water detector for muon flux measurement.
\end{abstract}

\maketitle


\section{\label{sec:Intro} Introduction}
The China Jinping Underground Laboratory (CJPL), located in Sichuan Province, China, is the world's deepest underground laboratory, with a vertical rock overburden of approximately 2400~m~\cite{Cheng:2017_CJPL_Review, Kang:2010_CJPL_Status, Ma:2021uzi}.
The 2400~m rock overburden at CJPL corresponds to a water-equivalent depth of 6720~m, making it deeper than other major underground laboratories, such as Gran Sasso (3800~m.w.e.)~\cite{Bellini_2012_Borexino_Muon}, Kamioka (2700~m.w.e.)~\cite{Eguchi_2003_KamLAND}, and SNOLAB (6000~m.w.e.)~\cite{Aharmim_2009_SNO_Muon}. 
This exceptional depth suppresses the cosmic-ray muon flux by approximately eight orders of magnitude relative to the flux at sea level, thereby providing an extremely low-background environment for rare-event experiments such as dark matter searches, neutrinoless double-beta decay, and neutrino physics~\cite{CDEX:2014amu, PandaX:2014mem, Chen:2017_PandaXIII, Jinping:2016iiq, Zeng:2020_Cosmogenic_Activation}.

\begin{figure*}[t]
\centering
\includegraphics[width=1.0\textwidth]{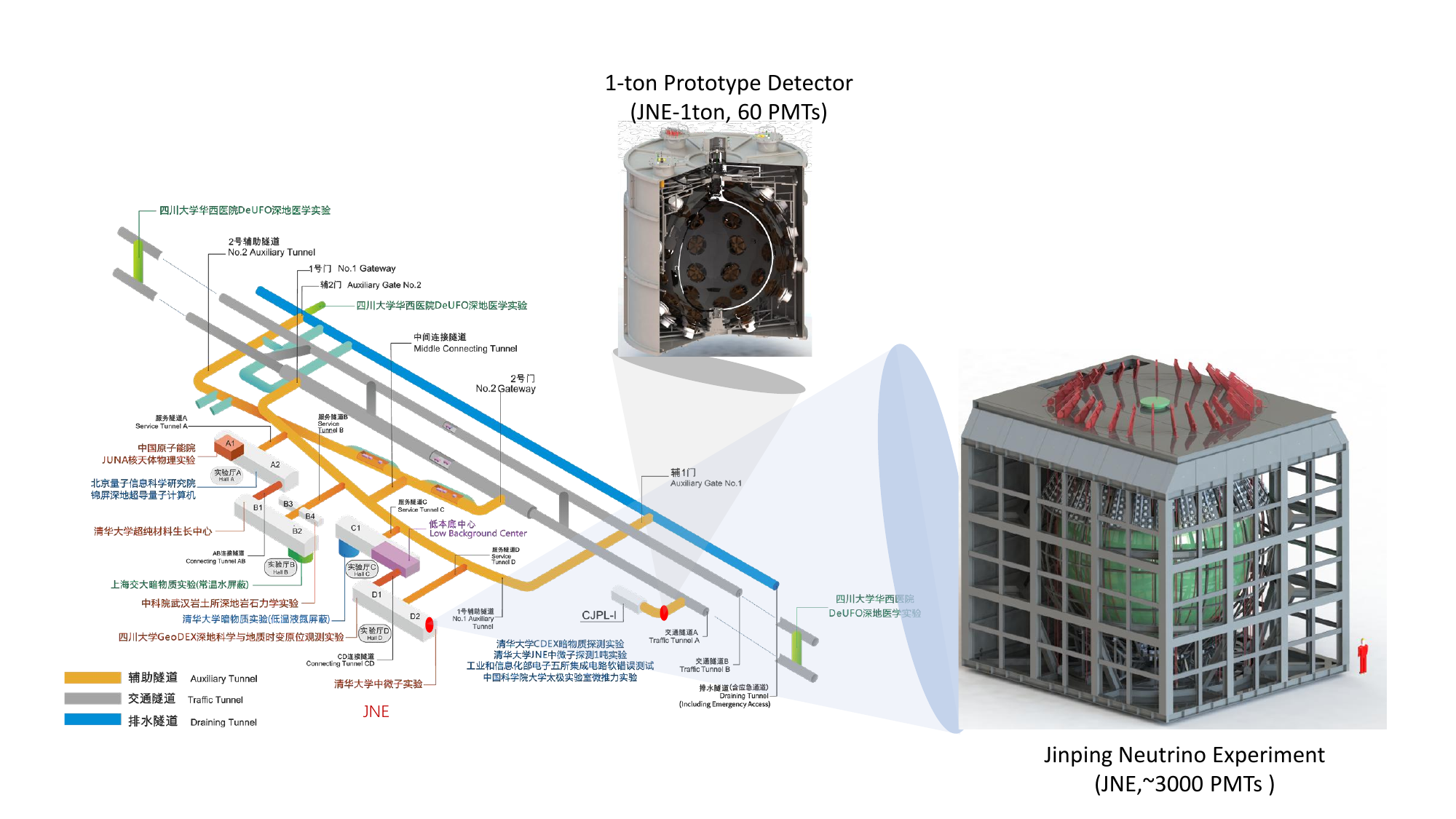}
\caption{Overview of the China Jinping Underground Laboratory. The JNE-1ton is located at CJPL-I, and the JNE detector is under construction.\cite{yang2026researchdevelopmentnewelectronics}}
\label{fig:CJPL_overview}
\end{figure*}

The Jinping Neutrino Experiment (JNE), located at CJPL-II(FIG.~\ref{fig:CJPL_overview}), aims to study MeV-scale neutrinos, including solar neutrinos, geoneutrinos~\cite{Sramek:2016_Jinping_Mantle, wan2017geoneutrinos}, and supernova relic neutrinos~\cite{wei2017discovery}; measurements of these neutrinos are highly sensitive to muon-induced backgrounds~\cite{Jinping:2016iiq}. Its 1-ton prototype detector, JNE-1ton, located at CJPL-I, is designed not only to validate the technical feasibility of JNE but also to provide several key measurement parameters for the design and optimization of the full-scale detector.

Cosmic-ray muons and the muon-induced backgrounds constitute an important background for neutrino studies: muons traversing the detector can mimic neutrino signals via direct energy deposition, while muon-induced spallation produces long-lived radioactive isotopes such as $^{11}$C, $^{9}$Li, and $^{8}$He~\cite{Galbiati:2005_C11,Li:2015_SK_Spallation,Agostini:2021_CNO_Evidence,Abe:2016_SKIV_Solar,Zhang_2024_Neutron_CJPLI}. A precise determination of the muon flux and its angular distribution is therefore essential for designing effective active and passive shielding strategies. The same muons can also be exploited as a probe: since their attenuation through rock depends on the density integrated along each line of sight, underground muon measurements can be used for mountain imaging, or muon radiography~\cite{Bonechi:2020muography,Li:2025MuographyChina,Zhang:2022_Muon_Tomography,JNE:2026muography}.

The cosmic-ray muon flux at CJPL has been measured by several experiments~\cite{Wu_2013_CJPLI_Muon}. At CJPL-I, the most recent measurement using the previous JNE-1ton liquid scintillator (LS) detector yields a muon flux of $(3.56 \pm 0.16_{\mathrm{stat}} \pm 0.10_{\mathrm{syst}}) \times 10^{-10}~\mathrm{cm}^{-2}\mathrm{s}^{-1}$, based on 1178 days of data and an average muon energy of 360~GeV \cite{Zhang_2024_Neutron_CJPLI}. 
At CJPL-II, a measurement with a plastic scintillator telescope over 1098 live days determined the flux to be $(3.03 \pm 0.24_{\mathrm{stat}} \pm 0.18_{\mathrm{sys}}) \times 10^{-10}~\mathrm{cm}^{-2}\mathrm{s}^{-1}$, the lowest value reported among underground laboratories worldwide \cite{Zhang_2025_CJPLII_Muon}.

In this work, we present a new measurement of the cosmic-ray muon flux at CJPL-I, obtained with the upgraded 1-ton prototype detector operating in water mode. 
This is the first muon flux measurement with a water-based detector at Jinping and significantly expands the available data sample~\cite{Guo_2021_CJPLI_Muon, Zhang_2024_Neutron_CJPLI}. Moreover, the water-mode operation provides an important validation of the water Cherenkov detection technique for future large-scale neutrino detectors at CJPL-II~\cite{Super-Kamiokande:2002weg, Jinping:2016iiq}.

This paper is organized as follows: Section~\ref{sec:1ton} describes the design and upgrades of the new 1-ton prototype detector; Section~\ref{sec:Evt} details the data analysis procedures, including dataset characterization, calibration, and event selection; the muon flux measurement, efficiency evaluation, and systematic uncertainty analysis are presented in the muon measurement section; and Section~\ref{sec:Rec} describes the template-based muon direction reconstruction method and its performance.

\section{\label{sec:1ton} The upgrade JNE-1ton detector}
The JNE-1ton is a small-scale prototype for the JNE, located in CJPL-I. Originally constructed in 2017~\cite{Wang:2017_1ton_Prototype}, the detector operated with slow LS~\cite{Guo:2017nnr} for six years and produced multiple physics results, including measurements of the cosmic-ray muon flux, cosmogenic neutron yield, and nature radiation background studies~\cite{Guo_2021_CJPLI_Muon, Zhang_2024_Neutron_CJPLI, Zhao:2022_Neutron_Yield, Zhang:2022_Muon_Tomography,wu2023performance}. 

The detector was decommissioned in September 2023 and subsequently underwent a comprehensive hardware upgrade. The upgraded detector serves as a verification platform for self-developed equipment and technologies destined for the full-scale JNE detector, and is capable of operating with both water and LS as the target material. 

The main goals of the 1-ton prototype are to validate the key detector technologies for the full-scale JNE experiment~\cite{wu2023performance}, including the slow liquid scintillator~\cite{Guo:2017nnr}, the 8-inch microchannel-plate photomultiplier tubes (MCP-PMT, GDB-6082)~\cite{Zhang:2023ued}, the readout electronics and self-developed waveform-digitizing systems~\cite{Yang:2024qqe, yang2026researchdevelopmentnewelectronics}, and the light concentrators~\cite{Ouyang:2025phk}. In addition, the prototype measures the underground background levels in situ, providing essential input for the background budget of the JNE experiment~\cite{wu2023performance}.

\subsection{Detector structure}
\begin{figure}[htbp]
\centering
\includegraphics[width=0.5\textwidth]{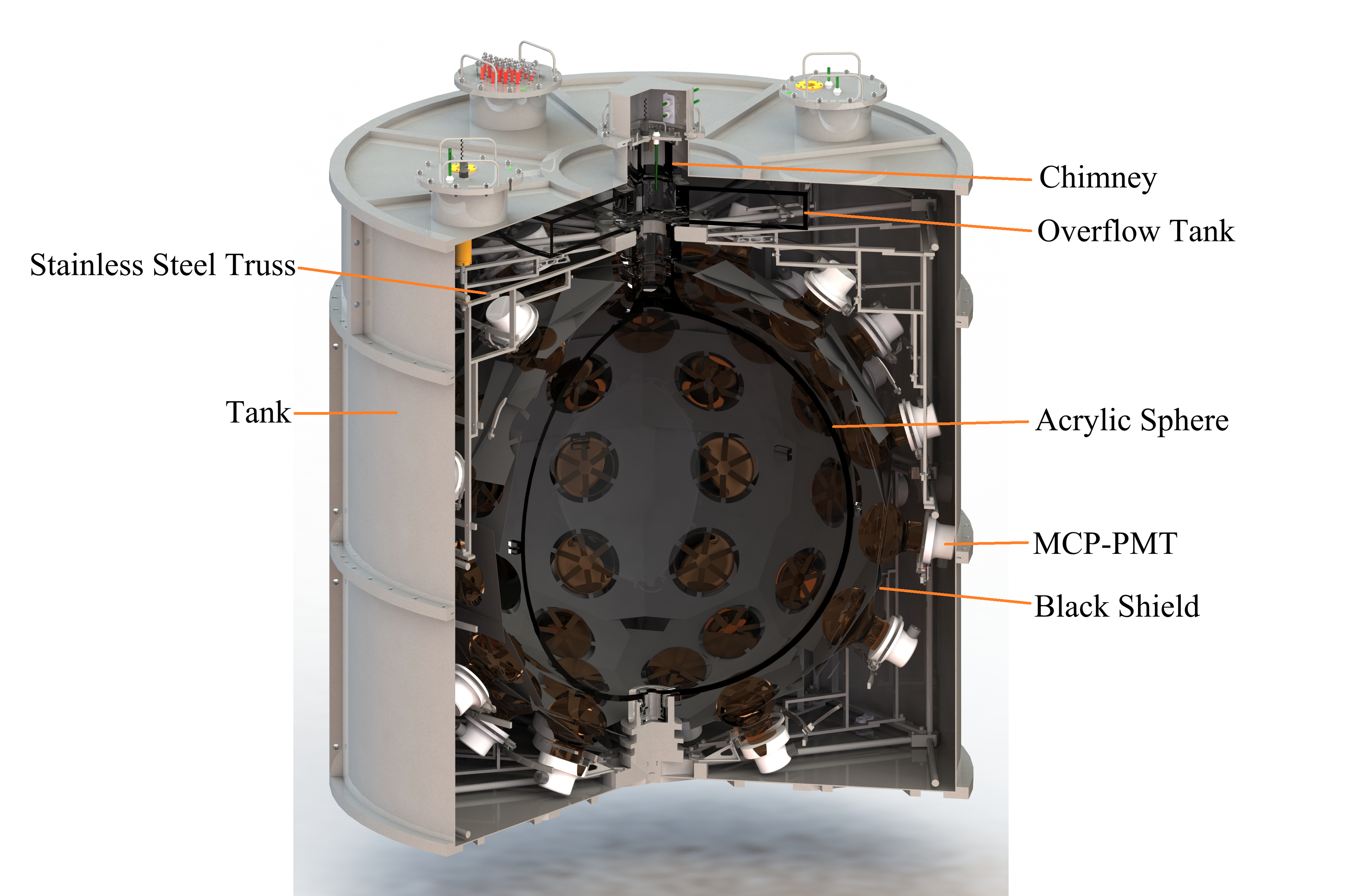}
\caption{Schematic diagram of the upgraded JNE-1ton detector. The central acrylic vessel is filled with water and surrounded by 60 MCP-PMTs. The black shield, installed during the Water-II phase, and the outer lead shielding wall are also indicated.}
\label{fig:detector_schematic}
\end{figure}

The detector maintains the same mechanical structure as the previous 1-ton prototype~\cite{Guo_2021_CJPLI_Muon,wu2023performance,Zhang_2024_Neutron_CJPLI}. 
The central target volume is defined by a spherical acrylic vessel with a radius of 0.645~m and a wall thickness of 20~mm, housed within a cylindrical stainless steel tank measuring 2000~mm in diameter and 2090~mm in height, as shown in FIG.~\ref{fig:detector_schematic}.

The new detector features a key upgrade in which the original 30 eight‑inch Hamamatsu R5912 PMTs are replaced by 60 eight‑inch MCP‑PMTs, doubling the photocathode coverage; this was enabled by a redesigned support truss that can accommodate the larger number of PMTs. The upgraded detector is divided into eight rings along the vertical direction. The first, second, seventh, and eighth rings from top to bottom each contain 5 PMTs, while the remaining rings each contain 10 PMTs, giving a total of 60. PMT numbering starts from the top first ring and proceeds counterclockwise.

The region between the acrylic vessel and the stainless steel tank wall is filled with water, serving as a passive shield against external gamma-ray and neutron backgrounds.

A significant new component of the upgraded detector is a black polyethylene (PE) shield installed around the PMT support structure during the Water-II phase, which was absent in the Water-I phase. 
This shield prevents most of the Cherenkov light produced in the water buffer region—outside the acrylic vessel but inside the steel tank—from reaching the PMTs.

The readout system employs eight CAEN V1751 Flash ADC boards and one CAEN V1495 logical trigger module, identical to the electronics configuration of the original detector~\cite{Guo_2021_CJPLI_Muon,wu2023performance}. Each V1751 board provides 8 channels with 10-bit ADC precision, a $\simeq$1~V dynamic range, and a 1~GSa/s sampling rate. All PMT signals are directly digitized by the V1751 boards.

In addition to the CAEN system, self-developed waveform-digitizing electronics have been tested on the detector during the LS phase~\cite{yang2026researchdevelopmentnewelectronics} to meet the future requirements of the full-scale JNE detector.

\subsection{Operational phases}
The upgraded detector has been operated through several data-taking periods with different target materials (water and LS) and detector configurations at CJPL-I. 

The data trigger logic is divided into two categories: single-PMT trigger and multi-PMT trigger. For a single-PMT trigger, the signal amplitude is required to be 5 mV below the baseline, with the baseline value independently calibrated before each data-taking run. The threshold for the multi-PMT trigger is set according to the detector configuration of each data-taking period: in the water mode, at least 10 PMTs and at least 4 PMTs must be triggered simultaneously in Water-I and Water-II, respectively, while in the LS mode the requirement is 45 PMTs.

The operational timeline is summarized as follows：
\begin{itemize}
\item Water-I took place from October 15, 2024, to January 8, 2025, with the acrylic vessel filled with water and without the black shield installed.
\item LS-I ran from January to April 2025, during which the detector was filled with LS using a water displacement method.
\item Water-II occurred from July 8 to September 25, 2025, with the black shield installed and water again serving as the target.
\item LS-II extended from September 2025 to January 2026, including energy calibration with an AmBe source, time calibration with an LED, and testing of the self-developed electronics.
\end{itemize}

This work focuses on the two water-phase datasets collected at CJPL-I: Water-I and Water-II, which were operated for 85 days and 79 days, respectively, corresponding to effective live times of 75.25 days and 69.88 days after the good-run selection (Sec.~\ref{sec:Evt}).

During Water-I, nine PMTs malfunctioned; after subsequent maintenance and upgrades, PMT reliability improved, and only four PMTs malfunctioned during Water-II. 
After completing its physics program at CJPL-I, the detector has been relocated to CJPL-II for future measurements.

\section{\label{sec:Evt} Data analysis}

\subsection{Datasets}
The data were collected during two water-phase periods at CJPL-I: Water-I and Water-II. The readout data are stored as TTree objects in ROOT files, each approximately 200~MB in size and containing about 15,000 events. Since Water-I and Water-II were triggered by at least 10 and 4 PMTs, respectively, the corresponding trigger rates are 31~Hz and 378~Hz. The vast majority of the events are low-energy radioactive background, and the difference in trigger conditions does not affect the muon detection. Owing to the different detector configurations in the two periods, the two datasets are processed separately.

\subsection{Good-run selection}
Some channels exhibit baseline instability, frequent discharge, or PMT malfunction, necessitating data quality selection. For each file, the baseline of every channel is calculated as the mean ADC value in the 20--120 ns range of the 980 ns waveform. A channel is flagged as bad if it satisfies at least one of the following criteria~\cite{Zhang_2024_Neutron_CJPLI}:
\begin{itemize}
\item The baseline variance exceeds 0.5~mV;
\item The mean baseline deviates from the overall mean baseline of that channel across all files by more than $5\sigma$;
\item The channel occupancy is more than $5\sigma$ above the average of all channels, or falls below 1\% of the average.
\end{itemize}

Files containing more than 4 bad channels (excluding those that are known to be faulty) are rejected. 

The reduction in live time stems mainly from two sources. First, part of each data-taking period was devoted to detector maintenance, commissioning, and dedicated calibration runs (e.g., the LED-based time-calibration measurements in Water-II); these intervals are not treated as physics runs. Second, the detector conditions were not fully stable at the beginning of each period: several channels exhibited unstable baselines or frequent discharges, and the PMT dark noise rates were elevated initially and decreased only after the detector had stabilized. Files recorded during these unstable periods are rejected by the good-run selection, which further reduces the effective live time.

After excluding the detector maintenance and test periods and applying the good-run selection, the live times for the two periods are $T_{\mathrm{Live,I}} = 75.25 \,~\mathrm{days}$ and $T_{\mathrm{Live,II}} = 69.88 \,~\mathrm{days}$, corresponding to good-run fractions of 88.53\% and 88.45\%, respectively.


\subsection{PMT dark noise}
The dark noise rate of each PMT is determined from he trigger window before main pulse in physics run, following the approach of the previous LS analysis~\cite{Zhang_2024_Neutron_CJPLI}. The procedure is as follows:
\begin{itemize}
    \item Dark noise hits are searched for in the 20--80~ns range before the main pulse;
    \item A threshold of 5 ADC bins is applied;
    \item The minimum time separation from the preceding event is $30\,\mu\mathrm{s}$, excluding after pulse;
    \item Only a single peak is permitted within the search window.
\end{itemize}

\begin{figure}[htbp]
\centering
\includegraphics[width=0.48\textwidth]{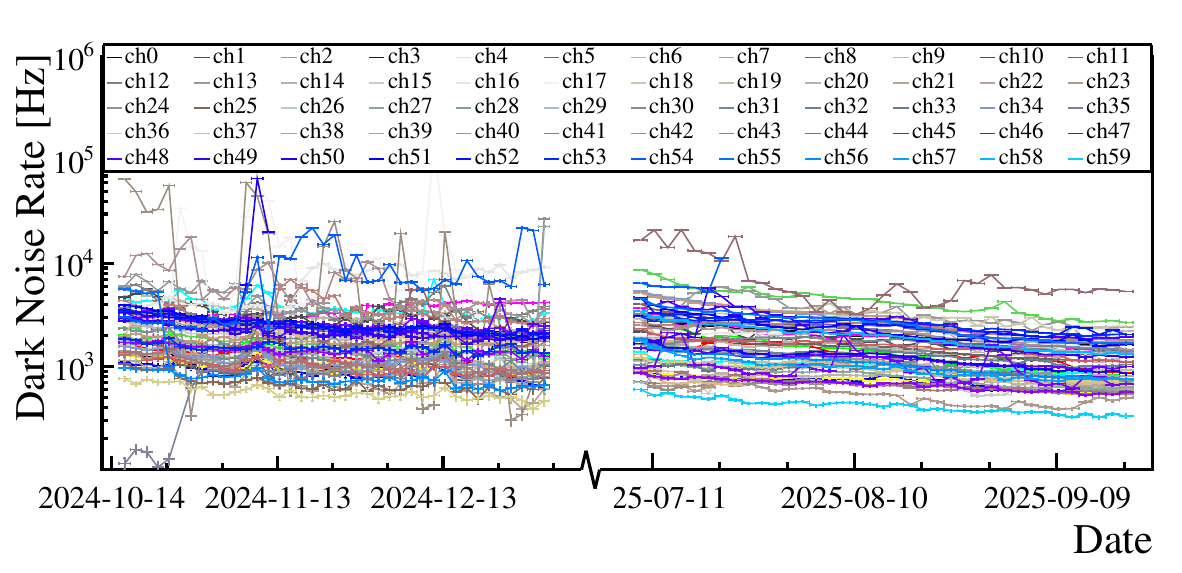}
\caption{Evolution of PMT dark noise rate on the Water-I and Water-II phases.}
\label{fig:darknoise}
\end{figure}

The resulting dark noise rate distribution, shown in FIG.~\ref{fig:darknoise}, exhibits a clear decreasing trend over time, reaching 1.1~kHz by September 2025. The measured dark noise rates serve as input parameters for the detector simulation and are used in PMT gain calibration.

\subsection{Gain calibration}
\begin{figure}[htbp]
\centering
\includegraphics[width=\linewidth]{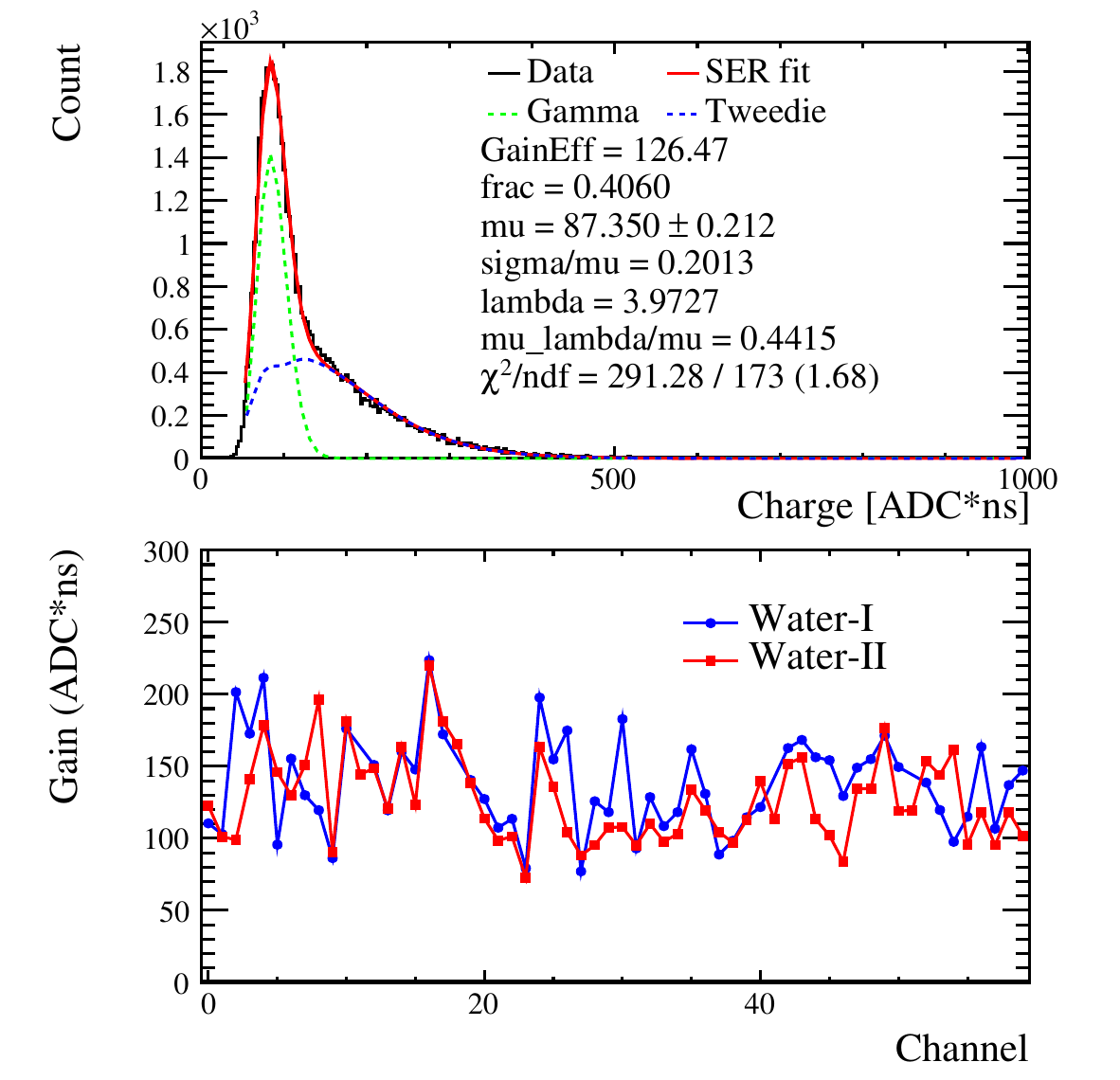}
\caption{Top: an example of SPE spectrum fit. The spectrum is from ch0 in Water-II. Bottom: PMT gain calibration result from dark noise charge spectrum fitting. It is worth noting that the primary difference between Water-I and Water-II lies in the replacement of several PMTs.}
\label{fig:gaincali}
\label{fig:fit_example}
\end{figure}

The gain of each PMT is determined by fitting the charge spectrum of its dark noise, which is taken as the single-photoelectron (SPE) spectrum, as illustrated in FIG.~\ref{fig:gaincali}. 
For MCP-PMTs, the SPE charge spectrum exhibits a characteristic long tail arising from the microchannel plate multiplication process~\cite{Zhang:2023ued}. 
The integrated charge distribution of SPE can be described by a Gamma + Tweedie function, from which the PMT gain is derived~\cite{Weng:2024tjs}.

\subsection{Photoelectron calculation}
The number of photoelectrons (PE) for each PMT channel is obtained by integrating the baseline-subtracted waveform over a fixed time window and dividing the resulting charge by the calibrated single-photoelectron charge of that channel. The same PE calculation procedure is applied to both data and simulation.

\begin{figure}[htbp]
\centering
\includegraphics[width=\linewidth]{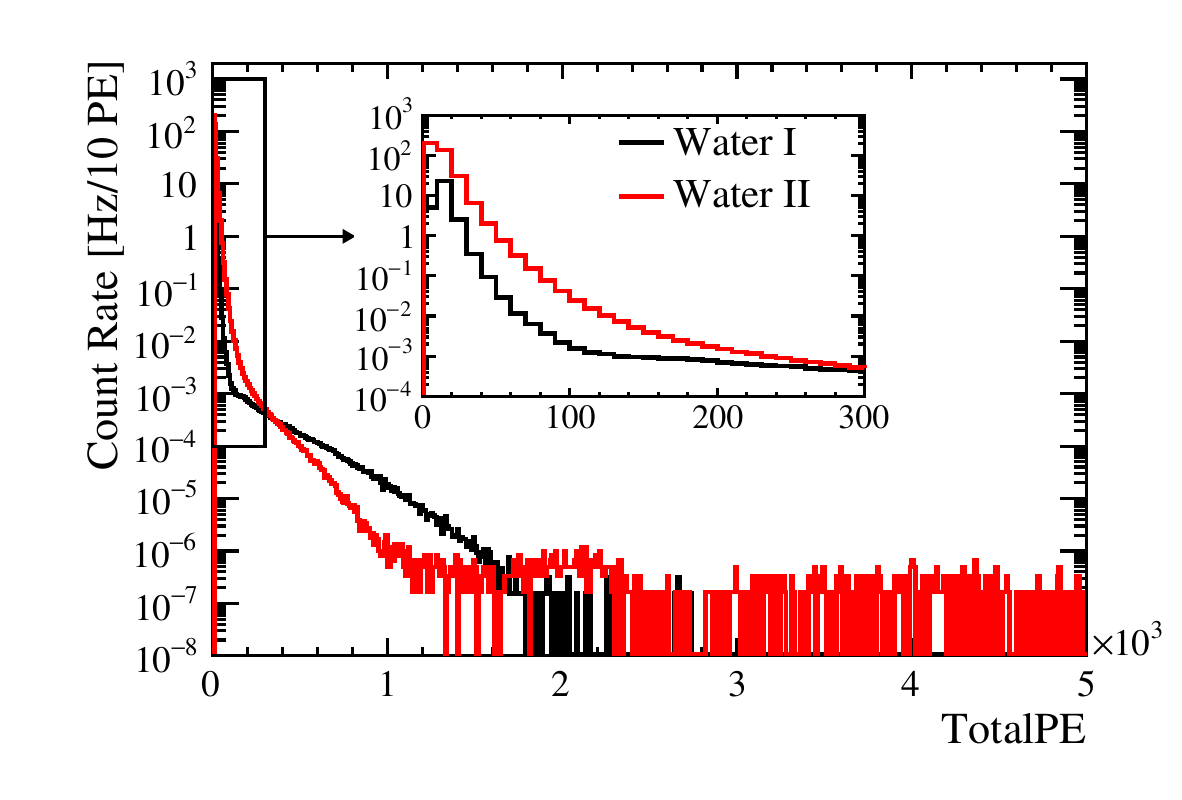}
\caption{Total PE spectra of the Water-I and Water-II datasets. The two periods were recorded with multi-PMT trigger requirements of at least 10 and 4 triggered PMTs, respectively.
}
\label{fig:PEspectrum}
\end{figure}

FIG.~\ref{fig:PEspectrum} shows the total-PE spectra of the Water-I and Water-II datasets, in which each event is represented by the total number of PEs summed over all channels. The two periods were recorded with different multi-PMT trigger requirements: at least 10 PMTs were required to be triggered simultaneously in Water-I, whereas at least 4 PMTs sufficed in Water-II (Sec.~\ref{sec:1ton}). 

Owing to the looser trigger condition, the counting rate of Water-II is higher than that of Water-I, most evidently at low total PE, where the spectra are dominated by radioactive background. 

In the high-energy region above about 500 PE, the remaining events are mainly PMT flashers as well as genuine muons; the flashers are removed by the flasher-rejection criteria of Sec.~\ref{sec:flasher-rejection}, and the surviving muon candidates are selected as described in Sec.~\ref{sec:Muon-selection}. 

Because the waveform digitizers have a finite dynamic range ($\simeq$ 1 V for CAEN V1751), some channels can saturate in high-energy events. In this analysis, only the charge within the measurable range is retained; the saturated portion is neither extrapolated nor compensated. The same treatment is applied in the simulation to ensure consistency between data and Monte Carlo for the energy-scale calibration and direction reconstruction.

\subsection{Flasher rejection}\label{sec:flasher-rejection}
\begin{figure}[htbp]
\centering
\includegraphics[width=\linewidth]{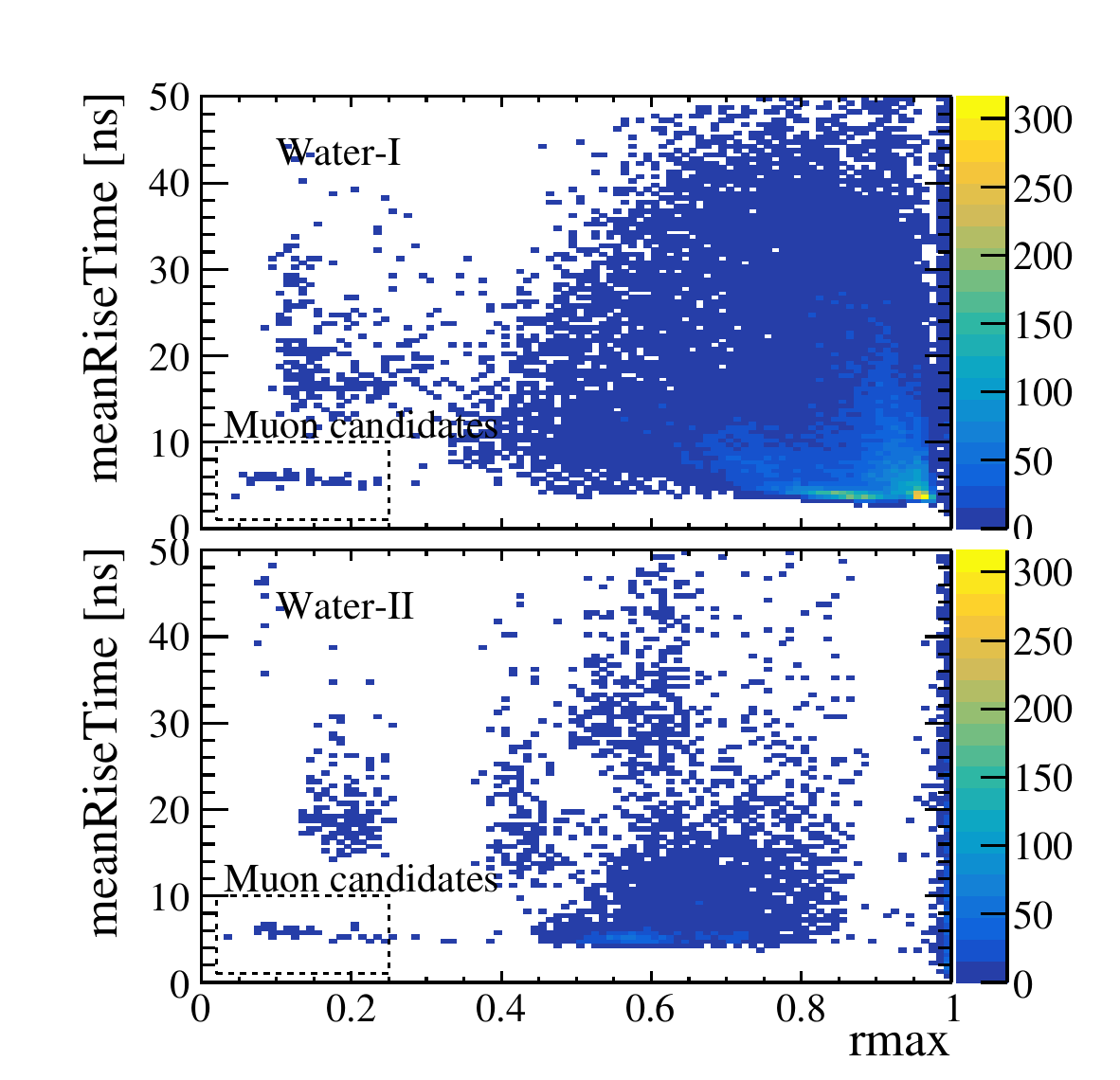}
\caption{Distribution of $r_\mathrm{max}$ versus $\mathrm{mrt}$ for muon candidate selection. Top: Water-I; bottom: Water-II. Flasher events are concentrated in the region $r_\mathrm{max} > 0.25$ and $\mathrm{mrt} > 10\,\mathrm{ns}$. The water-phase flasher distribution differs from that in the previous LS analysis~\cite{Zhang_2024_Neutron_CJPLI}. The presence of a continuously PMT flasher in Water-I led to poorer data quality compared with Water-II.
}
\label{fig:rmax_mrt}
\end{figure}

Background events with energy deposits comparable to muon signals arise primarily from PMT flashers. These can be effectively distinguished from genuine muon events using two discriminating variables: $r_\mathrm{max}$, defined as the ratio of the largest photoelectron (PE) count in a single channel to the total PE count of the event, and the mean rise time ($\mathrm{mrt}$), defined as the average rise time over all triggered channels. Flasher events are characterized by highly localized energy deposition and slow rise times, and are concentrated in the region $r_\mathrm{max} > 0.25$ and $\mathrm{mrt} > 10\,\mathrm{ns}$, as shown in FIG.~\ref{fig:rmax_mrt}. Flashers originating from different PMTs exhibit distinct distributions in the $r_\mathrm{max}$--$\mathrm{mrt}$ plane, possibly because the discharge occurs at different locations within the PMT. Events satisfying $r_\mathrm{max} < 0.25$ and $\mathrm{mrt} < 30\,\mathrm{ns}$ are retained as muon candidates.

\subsection{Time calibration}
Two complementary methods are used for the time calibration of 60 channels, and their results have been verified to be consistent on JNE-1ton.
The first method exploits the natural radioactive background in the detector: events originating from the central region are selected, and the PMT hit times are aligned to correct for intrinsic time offsets among electronics channels~\cite{wu2023performance}. 
The second method uses a 415~nm LED with a diffusion ball positioned at the detector center to perform a dedicated timing scan~\cite{yang2026researchdevelopmentnewelectronics}.

\begin{figure}[htbp]
\centering
\includegraphics[width=0.5\textwidth]{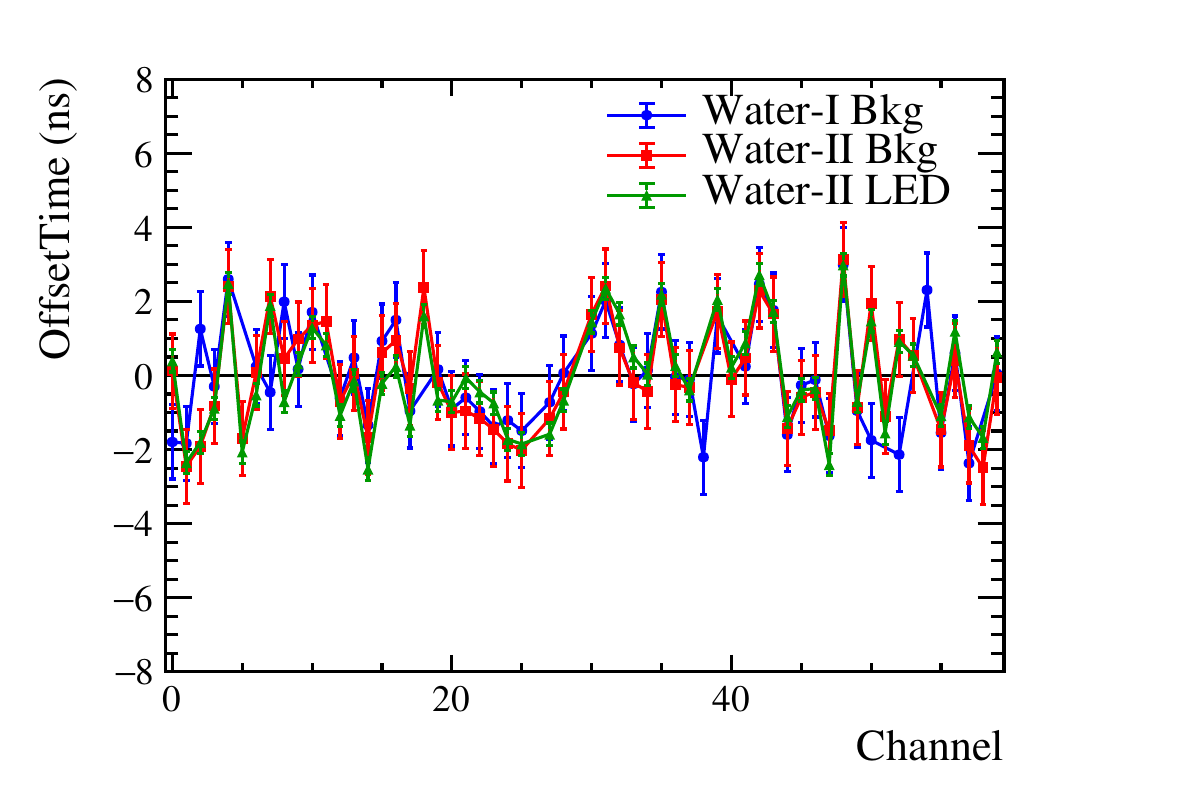}
\caption{Time calibration results obtained with the radioactive-background and LED methods. The two methods are assigned different uncertainties ($1.0\,\mathrm{ns}$ for the background method and $0.3\,\mathrm{ns}$ for the LED method), and the resulting calibration parameters are consistent with each other.}
\label{fig:timecali}
\end{figure}

For Water-II, the LED-based calibration yields a time calibration uncertainty of $0.3\,\mathrm{ns}$, as shown in FIG.~\ref{fig:timecali}; the dominant systematic uncertainty arises from the positioning of the diffusion ball, whose position uncertainty of 5~cm is conservatively translated into this value. For Water-I, where no LED scan was performed, the calibration is based on the environmental radioactive background. The dominant systematic uncertainty of this method comes from the non-uniformity of the radioactive background; based on a comparison with the LED calibration results in Water-II, an uncertainty of $1.0\,\mathrm{ns}$ is assigned to the time calibration.

\subsection{\label{sec:sim} Simulation}
The muon spectrum used in the simulation is taken from the mountain simulation of Ref.~\cite{Zhang_2024_Neutron_CJPLI}, following the previous JNE-1ton analyses~\cite{Guo_2021_CJPLI_Muon}. The detector simulation, however, uses the new geometry and PMT parameters of the upgraded detector, adapted to the actual operation configurations, including the 60-MCP-PMT array, the black shield, and the water buffer. This simulation is used for the efficiency evaluation and the template-based direction reconstruction presented below.

\begin{figure}[htbp]
\centering
\includegraphics[width=\linewidth]{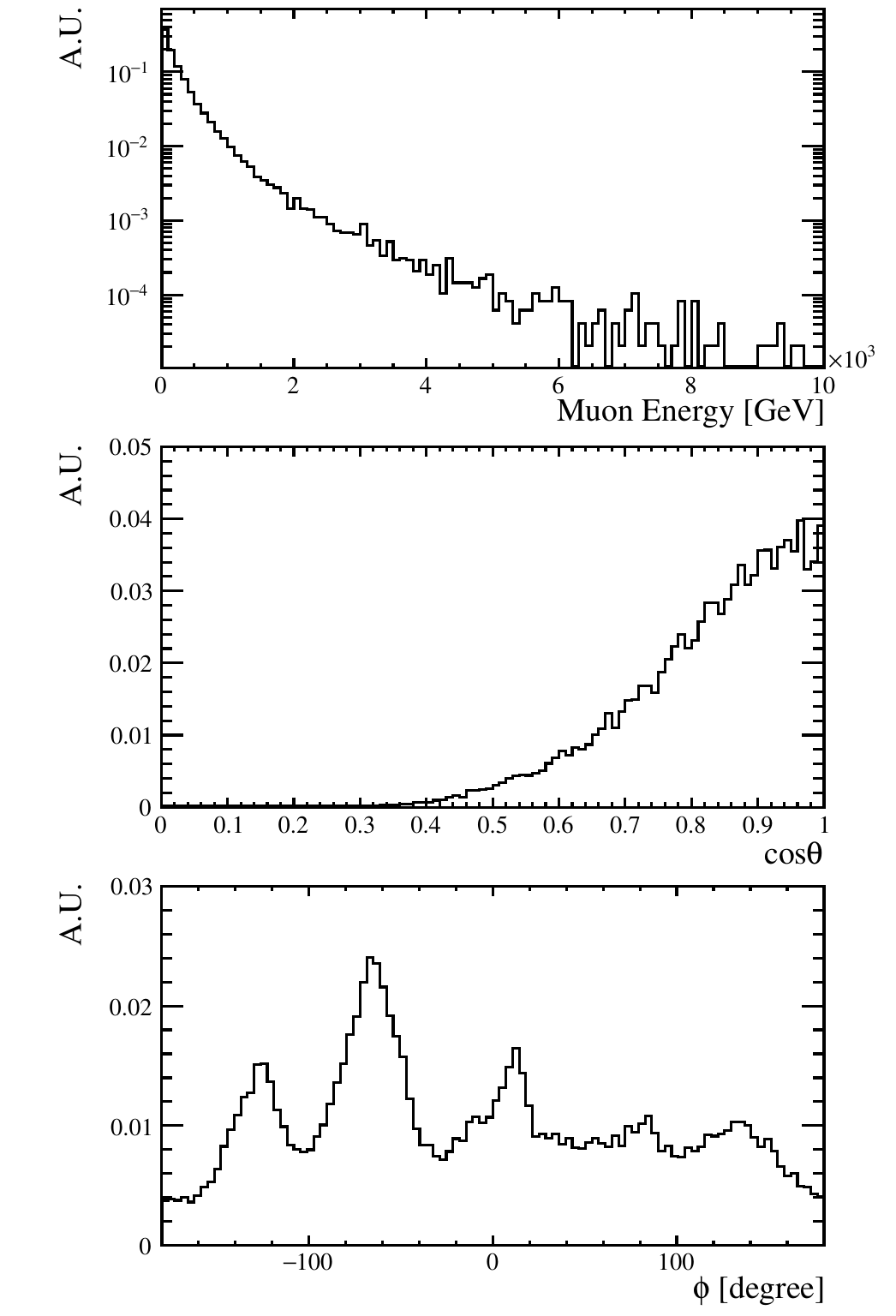}
\caption{Energy distribution and angular distribution of the muons generated by the mountain simulation used as input to the detector simulation. Top: energy distribution, with mean energy of about 360 GeV; middle: $\cos\theta$ distribution, dominated by small zenith angles; bottom: azimuthal $\phi$ distribution, whose structure reflects the azimuthal anisotropy of the mountain geometry at CJPL-I.}
\label{fig:muon_spectrum}
\end{figure}

The energy distribution and angular distribution of the muons generated by the mountain simulation is shown in FIG.~\ref{fig:muon_spectrum}. The top panel shows the energy distribution, with mean energy of about 360 GeV. The middle panel shows the $\cos\theta$ distribution, where the muon flux is concentrated at small zenith angles. The bottom panel shows the azimuthal $\phi$ distribution, whose pronounced structure originates from the azimuthal anisotropy of the mountain geometry.

\section{Muon flux measurement}

\subsection{Muon selection} \label{sec:Muon-selection}
The muon selection strategy differs between the two data-taking periods owing to the different detector configurations.

\begin{figure}[htbp]
\centering
\includegraphics[width=\linewidth]{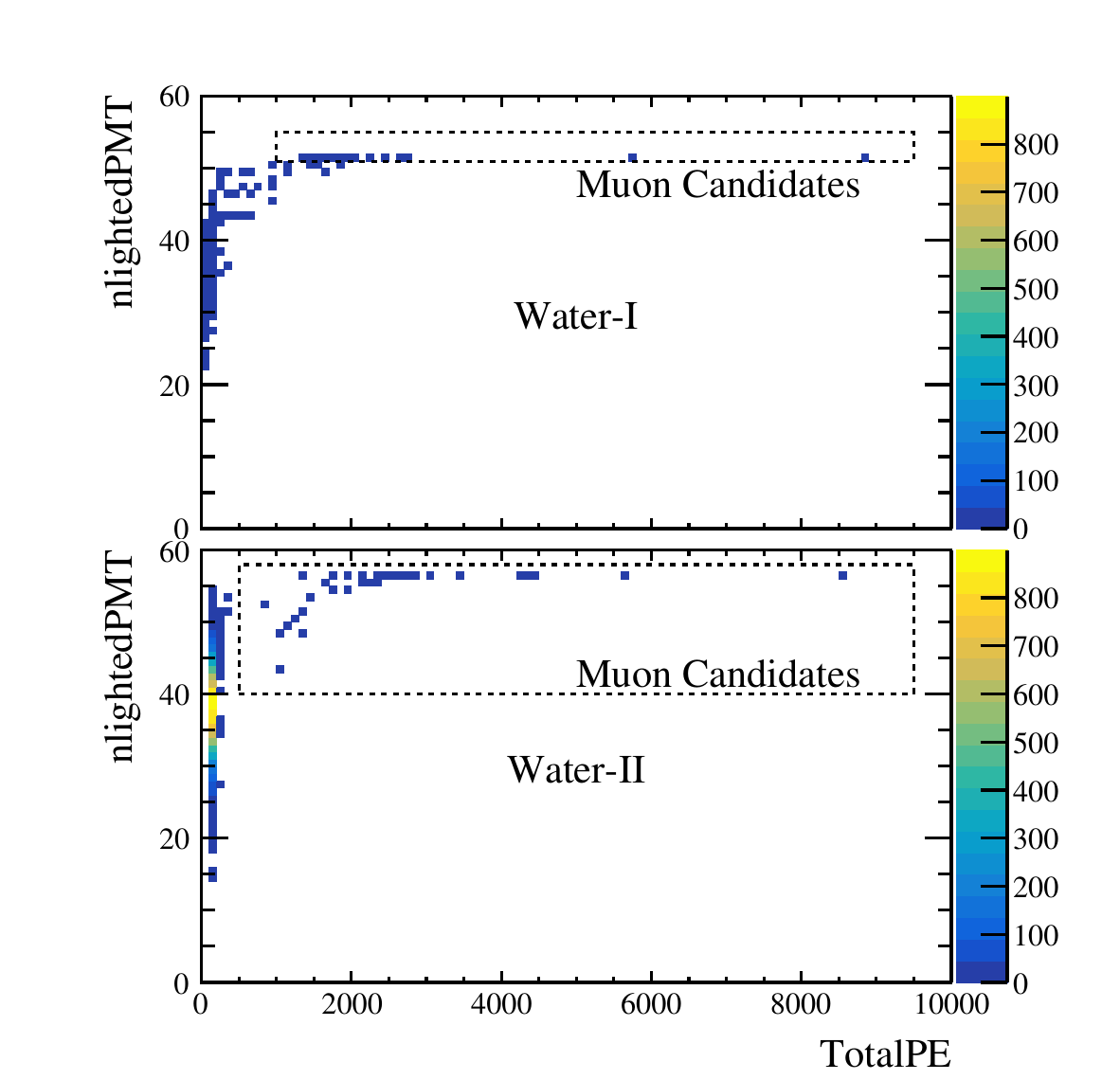}
\caption{Distributions of the total number of photoelectrons (PE) and the number of hit PMTs for muon candidates after flasher rejection. Top: Water-I; bottom: Water-II. The numbers of surviving PMTs are 51 and 56 for Water-I and Water-II, respectively.}
\label{fig:PEPMT1}
\end{figure}

In Water-I, where the black shield was absent, muons traversing the water in the steel tank produce Cherenkov light that can trigger the PMTs. These events are indistinguishable from the high-energy component of the radioactive background in terms of visible energy, but they exhibit a significantly larger number of triggered PMTs. Accordingly, events with at least 51 triggered PMTs (i.e., all surviving PMTs) are selected as muon candidates in Water-I, as shown in FIG.~\ref{fig:PEPMT1}, this criterion follows the previous LS analysis~\cite{Guo_2021_CJPLI_Muon,Zhang_2024_Neutron_CJPLI}.

\begin{figure}[htbp]
\centering
\includegraphics[width=1.0\linewidth]{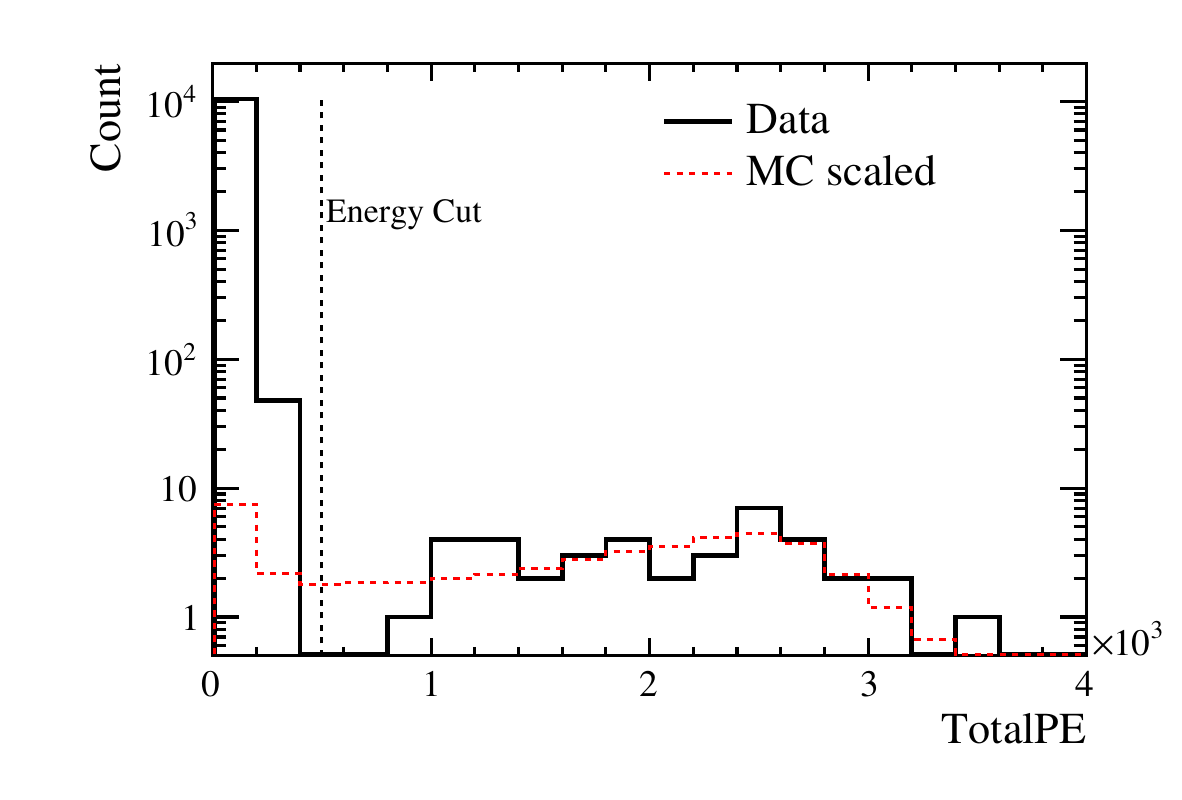}
\caption{Comparison of the visible energy spectra of muon candidates between Water-II data and simulation. An energy cut of more than 500 PE is applied for muon selection. Four events with total PE greater than 4000 are not shown.}
\label{fig:spectrum_data_mc}
\end{figure}

In Water-II, the black shield effectively blocks Cherenkov light originating from outside the detection region, leading to a clear separation between muon signals and background in the energy spectrum. Muon candidates in Water-II are therefore selected by requiring a visible energy greater than $500\,\mathrm{PE}$, as also illustrated in FIG.~\ref{fig:spectrum_data_mc}.

To justify the above selection criteria, we note that a 10~MeV beta particle in the simulation of either detector configuration can produce at most 48 triggered PMTs and about 400~PE. Therefore, the \(N_{\rm PMT}\ge 51\) cut in Water-I and the \(>500\)~PE cut in Water-II are both well above the radioactive-background level and can completely exclude it.

After applying all selection criteria, 26 muon events are retained from Water-I, while 45 muon events plus one additional special candidate are retained from Water-II. This special candidate is excluded from the flux calculation and is discussed separately in Sec.~\ref{sec:upgoing}.

\subsection{\label{sec:Flux} Muon flux measurement}
The muon flux for each data-taking period is calculated as
\begin{equation}
    \phi_{\mu,i}=\frac{N_{\mu,i}}{\varepsilon_i T_\mathrm{Live,i} S},
\end{equation}
where $N_{\mu,i}$ is the number of selected muon events, $\varepsilon_i$ is the muon detection efficiency and $T_{\mathrm{Live},i}$ is the effective data acquisition time for the $i$-th water mode, and $S$ is the equivalent projected area of the experimental hall. The latter is determined from the angular distribution of the underground muons~\cite{Guo_2021_CJPLI_Muon}; it depends only on the mountain topography above CJPL-I and is therefore common to the two water phases, with a value of $S = 78.5~\mathrm{m^2}$.

\subsubsection{Efficiency estimate}
The detection efficiency $\varepsilon$ is evaluated by the simulation, which applies the muon selection criteria to simulated muon events and computes the fraction of generated muons that survive all selection cuts, defined as:
\begin{equation}
    \varepsilon = \varepsilon_g \times \varepsilon_d + (1-\varepsilon_g) \times \varepsilon_o,
\end{equation}
where $\varepsilon_g$ is the fraction of muons arriving at the detector that pass through the acrylic vessel; $\varepsilon_d$ is the detection efficiency for these muons; $\varepsilon_o$ is the detection efficiency for the remaining muons that do not pass through the acrylic vessel.
A summary of the efficiencies for the two water phases is given in TAB.~\ref{tab:efficiency}.

\begin{table}[htbp]
\centering
\caption{Summary of the efficiencies for the muon flux measurement in Water-I and Water-II, comparing with previous result for LS phase\cite{Zhang_2024_Neutron_CJPLI}.}
\label{tab:efficiency}
\begin{tabular}{lcccc}
\hline
\hline
Phase & $\varepsilon_g$ & $\varepsilon_d$ & $\varepsilon_o$ & $\varepsilon$ \\
\hline
Water-I  & 2.1\% & 56.3\% & 0.3\% & 1.5\% \\
Water-II & 2.1\% & 98.0\% & 0.6\% & 2.7\% \\
LS\cite{Zhang_2024_Neutron_CJPLI} & 2.1\%&82.1\%&0.03\%&1.7\%\\
\hline
\hline
\end{tabular}
\end{table}

\begin{figure}[htbp]
\centering
\includegraphics[width=\linewidth]{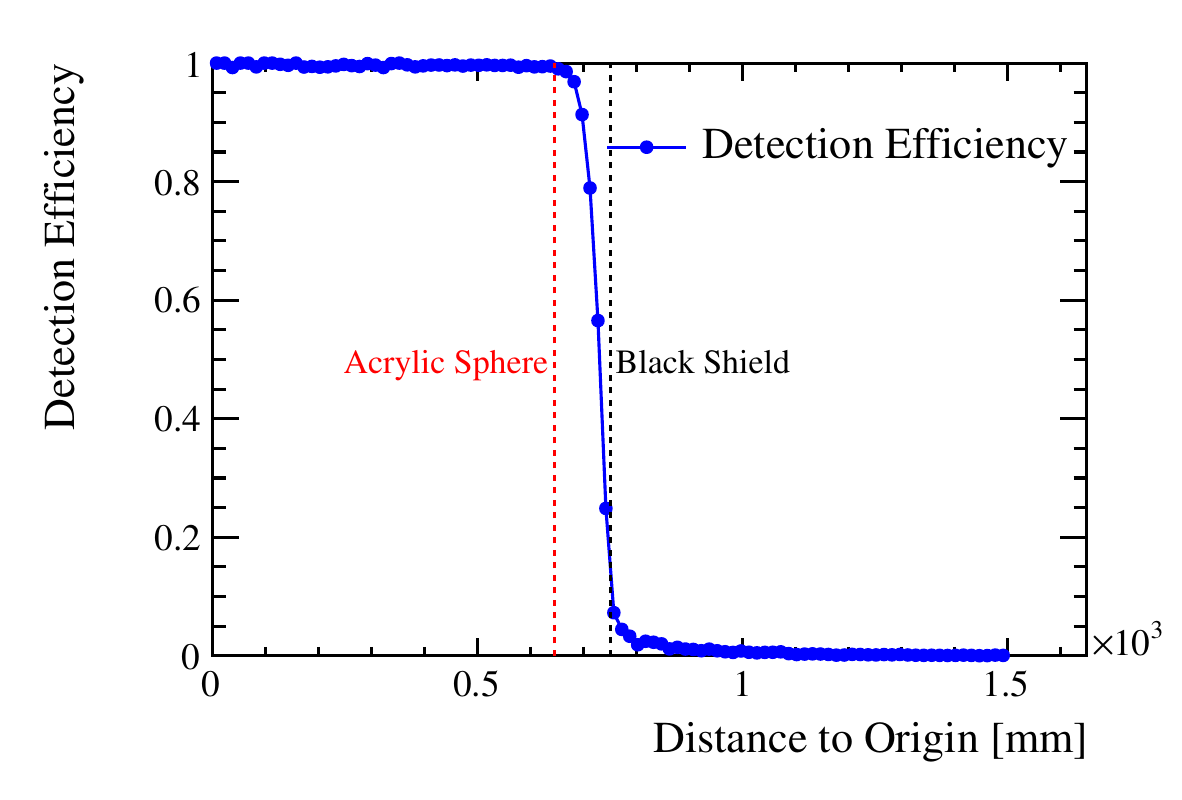}
\caption{Muon detection efficiency for muons crossing the JNE-1ton detector as a function of the distance from the center of the acrylic sphere to the muon track, obtained from the Water-II simulation. The radii of the acrylic sphere and the black shield are $0.65~\mathrm{m}$ and $0.75~\mathrm{m}$, respectively.}
\label{fig:efficiency_curve}
\end{figure}

The muon detection efficiency in Water-II is approximately 59\% higher than that reported for the previous LS JNE-1ton~\cite{Zhang_2024_Neutron_CJPLI}. This improvement originates mainly from the increase of the effective detection volume: in the water phases the sensitive volume is the water region enclosed by the black shield, which extends beyond the acrylic vessel, so that muons that miss the acrylic vessel but still traverse this region are detected with a much larger probability, whereas in the LS phase the sensitive volume is essentially the LS inside the acrylic sphere. The dependence of the detection efficiency on the distance between the muon track and the centre of the acrylic sphere, which reflects the enlarged effective detection volume, is shown in FIG.~\ref{fig:efficiency_curve}. 

\subsubsection{Systematic uncertainties}
The systematic uncertainties arise primarily from the detector simulation inputs and the data analysis procedures. A summary of all systematic uncertainty sources and their contributions to the flux measurement is presented in TAB.~\ref{tab:syst}.

\begin{table*}[htbp]
\centering
\caption{Summary of systematic uncertainty sources, their contributions to the muon flux measurement for Water-I and Water-II, and the correlation coefficient between them.}
\label{tab:syst}
\setlength{\tabcolsep}{12pt}
\begin{tabular*}{\textwidth}{@{\extracolsep{\fill}}lcccc}
\hline
\hline
\multirow{2}{*}{\textbf{Source}} & \multirow{2}{*}{\textbf{Uncertainty}} & \multicolumn{3}{c}{\textbf{Flux Uncertainty}} \\
                                 &                       & \textbf{Water-I}(Hit selection)       & \textbf{Water-II}(Energy selection)  & \text{Correlation}            \\
\hline
Latitude, Longitude        & ±100 m              & 1.2\%       & 2.7\% & 1\\
Elevation                  & ±100 m              & 2.0\%                    & 2.0\%       & 1        \\
Rock thickness             & ±50 cm              & 2.2\%                    & 0.6\%       & 1    \\
Physical model             & ±50\%               & 0.6\%                    & 1.3\%        & 1  \\
muon generator             &                     & 2.0\%                    & 1.7\%       & 1   \\
Acrylic sphere             & ±0.5 cm             & 0.3\%                    & 1.4\%       & 1            \\
PMT position               & {[}0.3, 2{]}cm      & 0.5\%                    & 1.3\%       & 0          \\
PMT angle                  & ±5 degree           & 1.2\%                    & 0.8\%       & 0        \\
PMT PDE &$\pm$ 10\%                   &    13.6\%        &   1.0\%            & 1 \\
Water absorption length    & {[}1/50, 1{]}       & 16.4\%             & 1.3\%             & 0  \\
Hole on shielding Sphere   & {[}0,10{]}mm        &                     & 3.4\%            & 0  \\
Radius of shielding Sphere & ±2 cm               &                     & 4.6\%            & 0 \\
Energy Scale               & ±10\%               &                     & 0.8\%            & 0\\
\hline
Total                      &                    &  21.7\% & 7.5\%                  & 0.16  \\   
\hline 
\hline
\end{tabular*}
\end{table*}

The individual sources of systematic uncertainty are described below. For the uncertainty sources that have already been studied in detail for the previous JNE-1ton, namely the geographic location and elevation, rock thickness, hadronic physics models, muon generator, and acrylic vessel radius, we adopt the same evaluation procedures as described in Refs.~\cite{Zhang_2024_Neutron_CJPLI, Farr:2007_SRTM, SK-water-absorption, Agostinelli:2003_Geant4, Guan:2015_Muon_Param, Fedynitch:2019_MCEq}.
Only the sources newly introduced by the water mode configuration are discussed in detail below.

\textbf{PMT position and angle:} The PMT positions and angles are newly introduced by the upgraded 60-MCP-PMT array and are varied within the mechanical installation tolerances of $[0.3, 2]$~cm and $\pm5^\circ$, respectively, affecting the solid-angle coverage and light collection efficiency.

\textbf{PMT PDE:}  The photodetection efficiency (PDE) directly affect the PE number detected by PMTs. In particular, when muon selection relies on the number of PMT hits, the PMTs at muon incidence positions with relatively few photoelectron counts are strongly affected, leading to a much larger uncertainty for Water‑I than for Water‑II. 

The PDE of the MCP-PMTs used in the simulation is taken from the factory test results of the PMTs. A relative uncertainty of 10\% is assigned to the PDE to account for the uncertainties of the factory measurements and the PMT-to-PMT variations. 

As only 10 PMTs were replaced between Water-I and Water-II, the uncertainty of PDE in two configurations is considered correlated.

\textbf{Water absorption length:} The attenuation of Cherenkov photons propagating through water is governed by the water absorption length. A shorter absorption length reduces the number of detectable photons, lowering the light collection efficiency and affecting both energy reconstruction and muon detection efficiency. 

The uncertainty is evaluated by varying the absorption length in the simulation by a scaling factor in the range $[1/50, 1]$ relative to the reference absorption length of Super-Kamiokande~\cite{SK-water-absorption}.

Since the water used in Water-I and Water-II was supplied by different equipments, the systematic uncertainties associated with water absorption in the two periods are tentatively treated as uncorrelated.

\textbf{Black shield tolerances (Water-II only):} In Water-II, the black shield introduces additional uncertainties from installation tolerances, including possible holes (up to 10~mm in diameter) and variations in the shield geometry ($\pm$2~cm), which affect the light collection geometry and the overall detection efficiency. 

\textbf{Energy scale (Water-II only):} The energy scale for muon events is calibrated using a binned $\chi^2$ minimization method in previous work\cite{Zhang_2024_Neutron_CJPLI}.However, because of the lack of a dedicated MeV-region energy calibration in Water-II, a more conservative 10\% uncertainty is adopted.

\subsubsection{Results}\label{sec:flux-results}
For the two individual data-taking periods, the measured muon fluxes are
\begin{equation}
\begin{aligned}
        \phi_{\text{I}}&=(3.60\pm0.71_{\mathrm{stat}}\pm0.78_{\mathrm{syst}})\times 10^{-10}\ \mathrm{cm}^{-2}\ \mathrm{s}^{-1}\\
        \phi_{\text{II}}&=(3.53\pm0.53_{\mathrm{stat}}\pm0.25_{\mathrm{syst}})\times 10^{-10}\ \mathrm{cm}^{-2}\ \mathrm{s}^{-1}
\end{aligned}
\end{equation}

Combining the measurements from the two data-taking periods, the muon flux obtained with the water-phase detector is
 \begin{equation}
 \phi_{\text{I+II}} = (3.55 \pm 0.43_{\mathrm{stat}}\pm 0.28_{\mathrm{syst}}) \times 10^{-10}~\mathrm{cm}^{-2}\mathrm{s}^{-1} 
\end{equation}

The combined analysis significantly reduces the statistical uncertainty, making it smaller than those of the individual Water-I and Water-II results. In contrast, because some systematic uncertainties are correlated between the two phases as illustrated in TAB.~\ref{tab:syst}, the improvement in the systematic uncertainty is limited, resulting in a combined systematic uncertainty that is slightly larger than that of the Water-II result alone.

This result is consistent with previous measurements at CJPL-I~\cite{Zhang_2024_Neutron_CJPLI}, confirming the reliability of the upgraded detector and the analysis methodology. The addition of the black shield markedly increases the accuracy of muon event identification, making it a crucial improvement in the Water‑II upgrade.

Since the Water-II configuration provides a better separation between muon signals and the radioactive background and yields a higher muon detection efficiency, the future water mode operation is planned to follow this configuration.

\section{\label{sec:Rec} Muon Direction Reconstruction}

The muon direction is reconstructed using a template-based method~\cite{Guo_2021_CJPLI_Muon}. Templates are generated from a detailed detector simulation by sampling muon entry positions and incident angles across the surface of the acrylic vessel, and recording the corresponding PMT hit times and PE distributions for each sampled trajectory. The incident directions are sampled uniformly over the full solid angle, so that the template library is omnidirectional and contains both downward- and upward-going trajectories; no directional prior is imposed, which is essential for the reconstruction of the up-going event discussed in Sec.~\ref{sec:upgoing}. For each muon event in data, its PMT hit pattern and PE distribution are compared against the template library, and the best-matching templates are identified using a distance metric, from which the muon entry position and incident direction are derived.

\subsection{JNE-1ton muon event characteristics}
In the water-phase operation, muon events are characterized by a clear Cherenkov light pattern and a high multiplicity of triggered PMTs.

\begin{figure*}[tbh]
\centering
\includegraphics[width=1.0\textwidth]{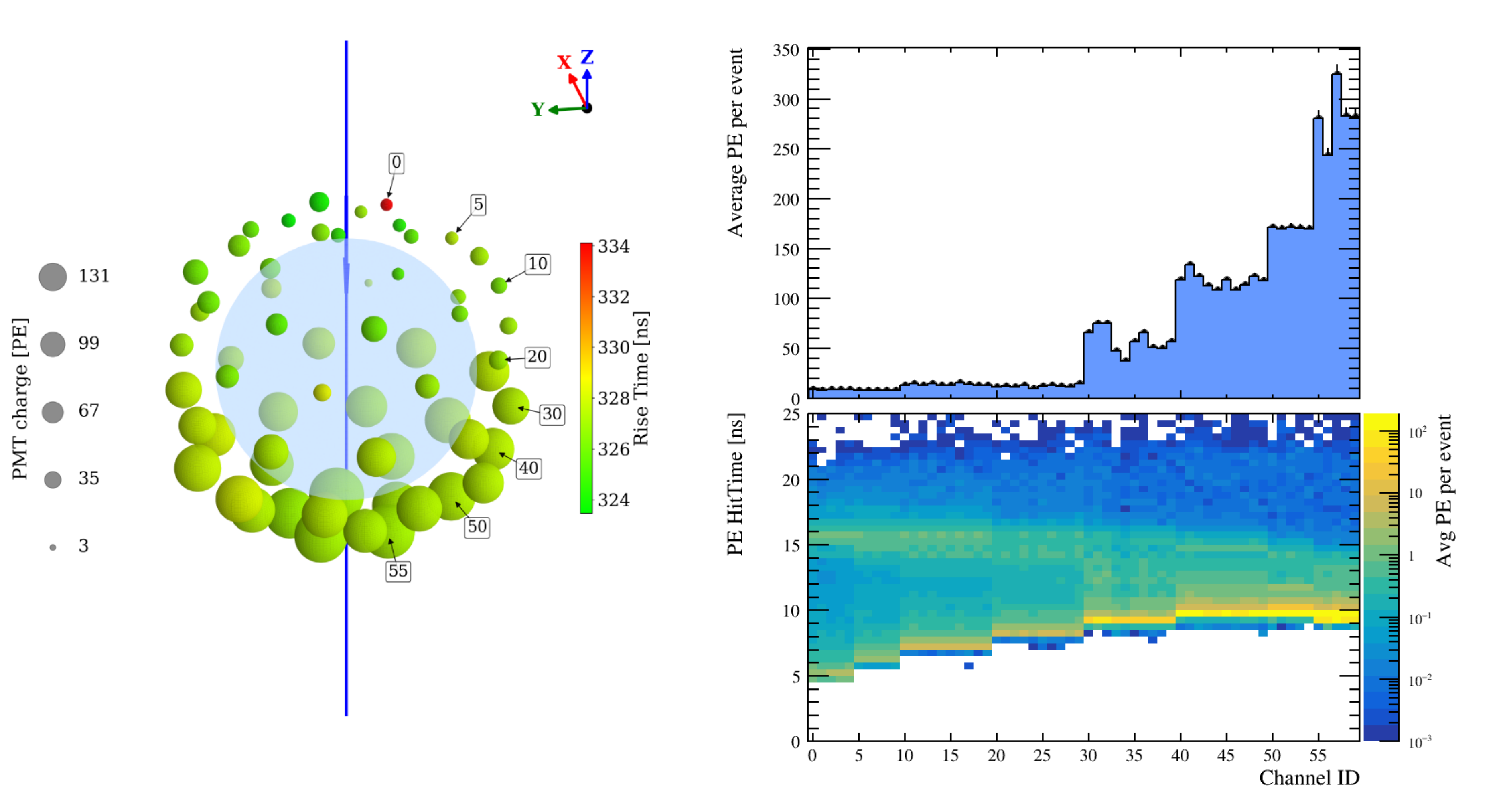}
\caption{Combined event display for a simulated muon incident from directly above. Left: schematic of the detector showing the Cherenkov hit pattern. The top PMTs are triggered first, while the bottom PMTs have the largest PE counts. The starting PMT number (Channel ID) of each ring is also indicated. Upper right: distribution of average PE per event, showing differences among PMTs at the same ring due to variations in PDE of PMT. Lower right: PE time distribution.}
\label{fig:event_disp_combined}
\end{figure*}

The spatial distribution of PE and the ordering of PMT hit times reflect the muon trajectory: PMTs close to the entry point are hit earlier, while the region with the largest PE concentration corresponds to the exit point or the track segment with the longest path in the active water volume.
These features provide the physical basis for the template-based direction reconstruction described below. The combined event display in FIG.~\ref{fig:event_disp_combined} shows a simulated muon incident from directly above: the top PMTs are lit first, while the bottom PMTs record the largest PE counts.

\subsection{Template method}
The template method has been optimized to extend its applicability to a wider range of detector configurations, including the water mode, and to improve the muon direction reconstruction precision. 
The key ingredient is a generalized distance that combines per-PMT time and PE information to compare each data event with simulated templates.

For two given events $m$ and $n$, the time and PE information of each channel is expressed as:
\begin{equation}
\begin{aligned}
    \mathbf{T}_m=\left(t_{m1},t_{m2},\cdots,t_{m_N}\right), \\
    \mathbf{Q}_m=\left(q_{m1},q_{m2},\cdots,q_{m_N}\right),
\end{aligned}
\end{equation}
where $N$ is the total number of valid channels, and $q_{mi}$ is the PE count in the $i$-th channel of event $m$.

Because the absolute trigger time of an event carries no direction information, each event is first aligned by removing its global time offset. For each event, the average PE time is defined as the PE-weighted mean of the channel trigger times:
\begin{equation}
    \bar{t}_m=\frac{\sum_{i=0}^N q_{mi}t_{mi}}{\sum_{i=0}^N q_{mi}}.
\end{equation}
The relative time of each channel is then obtained by subtracting this average:
\begin{equation}
    \bar{\mathbf{T}}_m=\left(t_{m1}-\bar{t}_m,t_{m2}-\bar{t}_m,\cdots,t_{m_N}-\bar{t}_m\right).
\end{equation}

The generalized distance between two events is defined as the sum over all channels of the squared differences in relative time and PE count, weighted by their inverse variances:
\begin{equation}
\begin{aligned}
    d^2(m,n)=\sum_{i=0}^{N}\left[d_Q^2(m,n,i)+d_T^2(m,n,i)\right],
\end{aligned}
\end{equation}

For each muon event, the $k$ templates with the smallest values of $d^2(m,n)$ are selected for the direction reconstruction. The method for deriving the direction from these $k$ nearest templates is the same as that in Ref.~\cite{Guo_2021_CJPLI_Muon}. The distance $d_Q^2(m,n,i)$ and $d_T^2(m,n,i)$ for per channel are given by
\begin{equation}
\begin{aligned}
    d_T^2(m,n,i)&=\left(\bar{T}_{mi}-\bar{T}_{ni}\right)^2\slash f(m,n,i),\\
    f(m,n,i)&=\left(\frac{T_s^2}{Q_{mi}}+\frac{T_s^2}{Q_{ni}}\right);\\
    d_Q^2(m,n,i)&=\left(Q_{mi}-Q_{ni}\right)^2\slash g(m,n,i),\\
    g(m,n,i)&=\left[\left(Q_{mi}+Q_{ni}\right)\left(1+\frac{\sigma_{\mathrm{SPE}}^2}{\mu_{\mathrm{SPE}}^2}\right)\right];
\end{aligned}
\end{equation}
where $\bar{T}_{mi}$ and $Q_{mi}$ denote the $i$-th components of $\bar{\mathbf{T}}_m$ and the PE-count vector $\mathbf{Q}_m$, respectively.  $T_s$ and $\left(1+\frac{\sigma_{\mathrm{SPE}}^2}{\mu_{\mathrm{SPE}}^2}\right)$ are event-independent constants: the former is determined by the optical response properties of the detector, and the latter by the single-PE response characteristics of the PMT.

When the template method is applied to experimental data, the uncertainties in time and gain calibration must also be accounted for. The weight functions are accordingly modified as:
\begin{equation}
\begin{aligned}
f(m,n,i)=&\frac{T_s^2}{Q_{mi}}+\frac{T_s^2}{Q_{ni}}+\sigma^2_{\mathrm{TCali}},\\
g(m,n,i)=&(Q_{mi}+Q_{ni}+Q_{mi}\eta^2)\left(1+\frac{\sigma_{\mathrm{SPE}}^2}{\mu_{\mathrm{SPE}}^2}\right),
\end{aligned}
\end{equation}
where $\sigma_{\mathrm{TCali}}$ is the time calibration uncertainty and $\eta$ is the relative gain calibration uncertainty.

A direct application of the above distance requires the triggered channels of the template and the muon event to be exactly identical. In the water-phase configuration, however, muon events typically produce a limited number of triggered channels with a non-uniform spatial distribution, rendering the direct template matching ineffective. To overcome this limitation, a marginal likelihood approach is adopted, which enables the comparison of events and templates with only partially overlapping sets of triggered channels, thereby extending the method's applicability to the water phase and other sparse-hit scenarios.

For a single channel in which one event registers zero PE while the other has a non-zero PE count $Q$, the marginal likelihood is constructed from the Poisson distribution as:
\begin{equation}
\begin{aligned}
\mathcal{L}(0|Q)&=\int_{0}^{+\infty}\sum_{n=0}^{\infty}\pi(0,k)\,\pi(n,k)\,P(Q,n)\ \mathrm{d} k\\
&=\sum_{n=0}^{\infty}\frac{P(Q,n)}{2^{n+1}},
\end{aligned}
\end{equation}
where $k$ is the expected number of PEs at the PMT, $\pi(n,k)$ is the Poisson probability of observing $n$ PEs given expectation $k$, and $P(Q,n)$ is the probability of measuring PE count $Q$ given $n$ PEs, calculable from the PMT SPE spectrum. Using the likelihood ratio, the corresponding $\chi^2$ contribution is:
\begin{equation}
d_Q^2=-2\ln\frac{\mathcal{L}(0|Q)}{\mathcal{L}(0|0)}. \label{chi2}
\end{equation}
\begin{figure}[htbp]
\centering
\includegraphics[width=\linewidth]{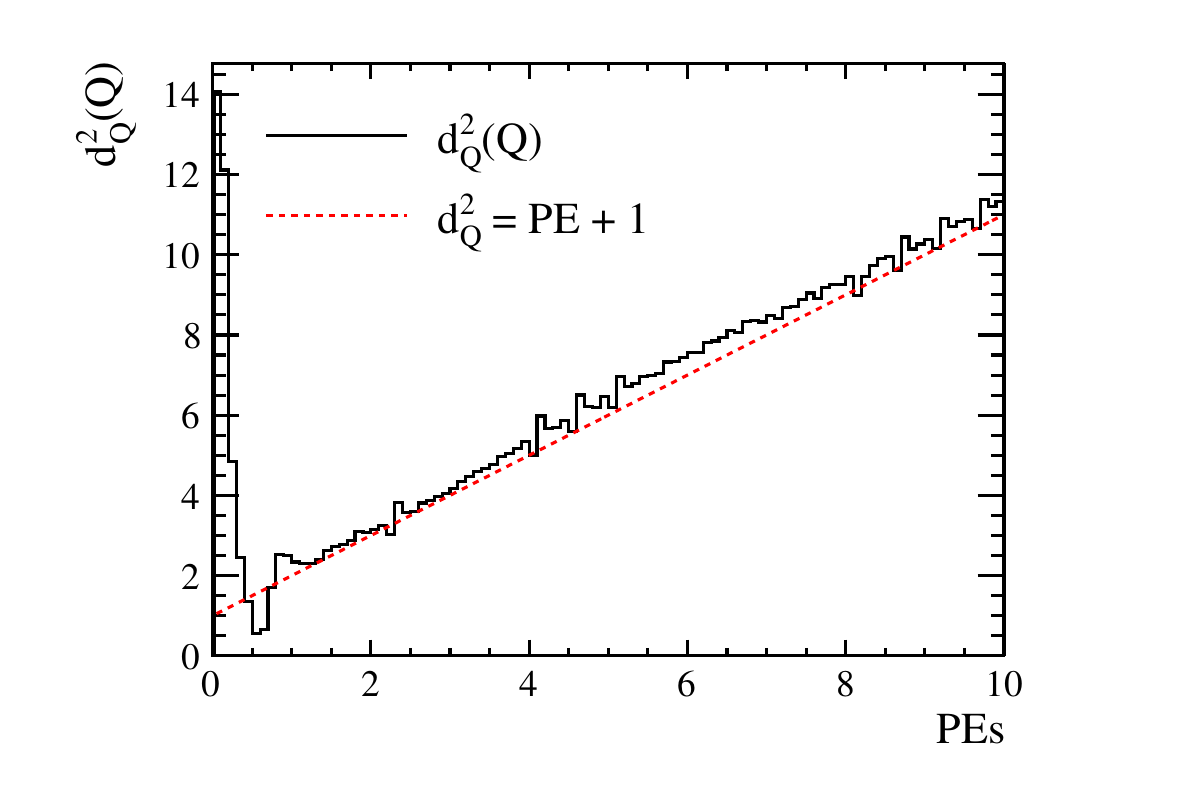}
\caption{The $d_Q^2$ calculated with Eq. \eqref{chi2}, using the typical SPE spectrum like FIG.~\ref{fig:fit_example}. }
\label{fig:Chi2Q}
\end{figure}

As shown in FIG.~\ref{fig:Chi2Q}, this $\chi^2$ contribution is well approximated by
\begin{equation}
d_Q^2\approx \mathrm{PE}+1.
\end{equation}

The resulting PE term is incorporated into the distance function, with the corresponding time term set to unity. To prevent excessively large reconstruction discrepancies, if the non-zero PE count exceeds $5$\,PE, the channel is deemed incompatible and the template is rejected.

\subsection{Angular resolution improvement}
The optimized reconstruction method significantly improves accuracy compared to the original approach and extends reliable reconstruction capability to a broader range of experimental configurations.

Throughout this work the angular resolution is defined as the 68.3\% quantile of the distribution of the included angle $\Delta\Theta$ between the reconstructed and the true muon direction in the Monte Carlo sample, i.e.\ 68.3\% of the reconstructed events have an angular error smaller than the quoted value; the same estimator is applied to all the methods compared below.

\begin{figure}[htbp]
\centering
\includegraphics[width=\linewidth]{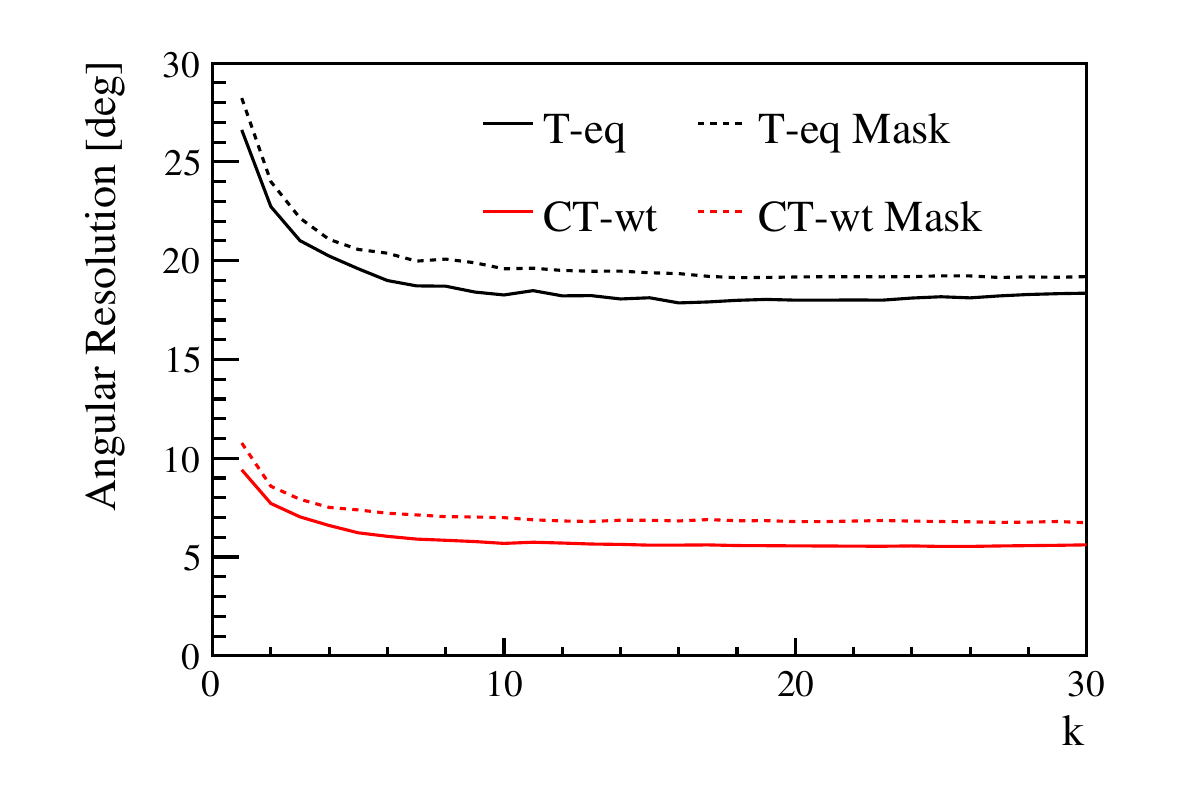}
\caption{Comparison of the muon direction reconstruction resolution between the time-only (T-eq) method of Guo et al.~\cite{Guo_2021_CJPLI_Muon} and the PE-time weighted (PT-wt) method of this work, using only events with all PMTs triggered (56 PMTs). The horizontal axis is the number \(k\) of nearest templates used in the reconstruction; the resolution gradually saturates for \(k \gtrsim 10\). The mask test, in which the 26 odd-numbered PMTs are ignored, illustrates the effect of the photocathode coverage on the resolution.}
\label{fig:resolution}
\end{figure}

To validate both the improved reconstruction method and its effectiveness for non-full-channel triggering, 10,000 Monte Carlo muon events and 140,000 reconstruction templates with Water-II configuration are used. FIG.~\ref{fig:resolution} compares the reconstructed angular resolution obtained with the time-only (T‑eq) method of Guo et al.~\cite{Guo_2021_CJPLI_Muon} and the PE‑time weighted (PT‑wt) method developed in this work, using only events with all PMTs triggered (56 PMTs). The horizontal axis is the number \(k\) of nearest templates used in the reconstruction; the angular resolution gradually saturates when \(k\) exceeds about 10. The comparison shows that the improved algorithm significantly enhances the resolution, owing largely to the inclusion of PE information. With the full 60-PMT array and the PE-time weighted method the 68.3\% containment angle is about 6$^{\circ}$. For the comparison with Ref.~\cite{Guo_2021_CJPLI_Muon}, whose result is quoted as an average $\Delta\Theta$, the corresponding mean of the $\Delta\Theta$ distribution is used, $5.9^{\circ}$; this is directly comparable with the average $\Delta\Theta \simeq 20^{\circ}$ obtained with the time-only method for the masked configuration that reproduces the previous 30-PMT coverage~\cite{Guo_2021_CJPLI_Muon}, so that the angular resolution is reduced to about 30\% of its previous value, i.e.\ by a factor of about 3.4, marking a major advance.

A mask test excluding the 26 odd‑numbered PMTs further indicates that the additional photocathode coverage also improves the resolution. Moreover, the angular resolution from the T‑eq mask test is close to that of the previous JNE‑1ton (30‑PMT) setup~\cite{Guo_2021_CJPLI_Muon}.

\begin{figure}[htbp]
\centering
\includegraphics[width=\linewidth]{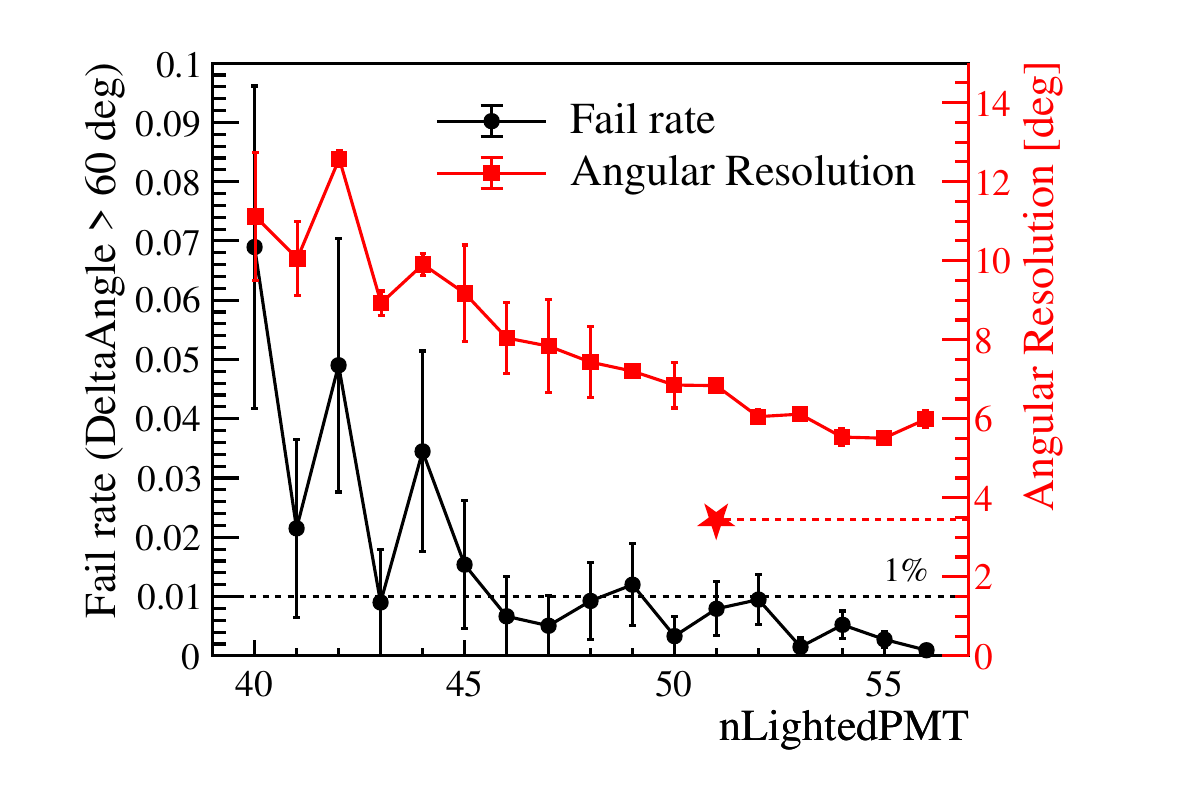}
\caption{Reconstruction failure rate (left axis) and angular resolution (right axis) as functions of the number of triggered PMTs, illustrating the reconstruction performance for events with incomplete channel triggering. The red star indicates the resolution of 3.44$^\circ$ for 51 PMT triggered events with Water-I configuration. In Water-I, the absence of the black shield makes the timing feature of the PMTs near the incident point more prominent, resulting in much better reconstruction precision than in Water-II. 
}
\label{fig:eff_res_vs_nPMT}
\end{figure}

FIG.~\ref{fig:eff_res_vs_nPMT} illustrates the reconstruction performance for events with incomplete channel triggering. The failure rate is kept at a low level, and the reconstruction precision remains close to that of the full-trigger case, demonstrating effective reconstruction even when not all PMTs are fired.

\subsection{Muon direction distribution}
\begin{figure}[htbp]
\centering
\includegraphics[width=\linewidth]{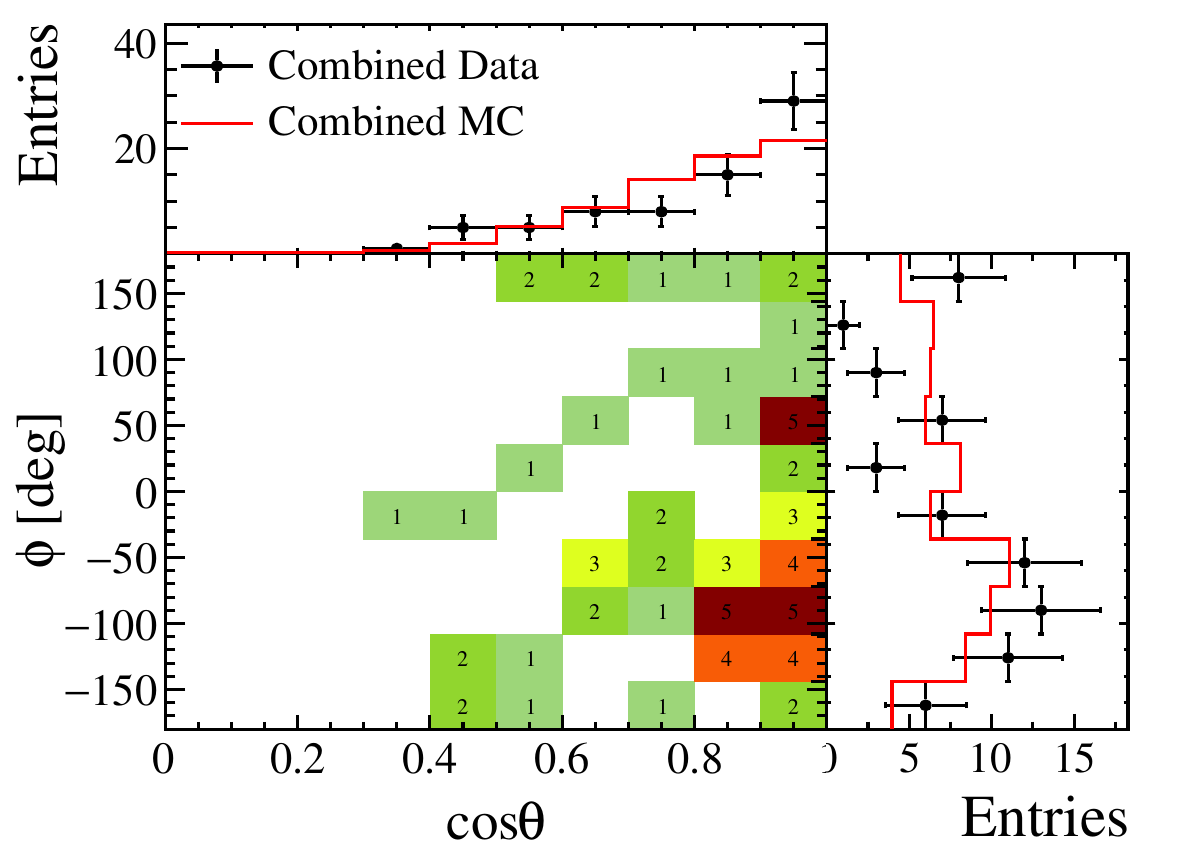}
\caption{Compared with the simulation results, the reconstructed zenith and azimuth angle distributions of the muon data candidates comprise 26 events for Water‑I and 45 events for Water‑II, respectively.}
\label{fig:angulardist}
\end{figure}

The final reconstruction results, as shown in FIG.~\ref{fig:angulardist}, demonstrate that the distributions of the reconstructed zenith and azimuth angles are in good agreement with the muon distribution predictions in the CJPL-I~\cite{Guo_2021_CJPLI_Muon,Zhang_2024_Neutron_CJPLI}, thereby validating the detector's direction reconstruction capability.

\subsection{Up-going muon event}
\label{sec:upgoing}
\begin{figure}[htbp]
\centering
\includegraphics[width=\linewidth]{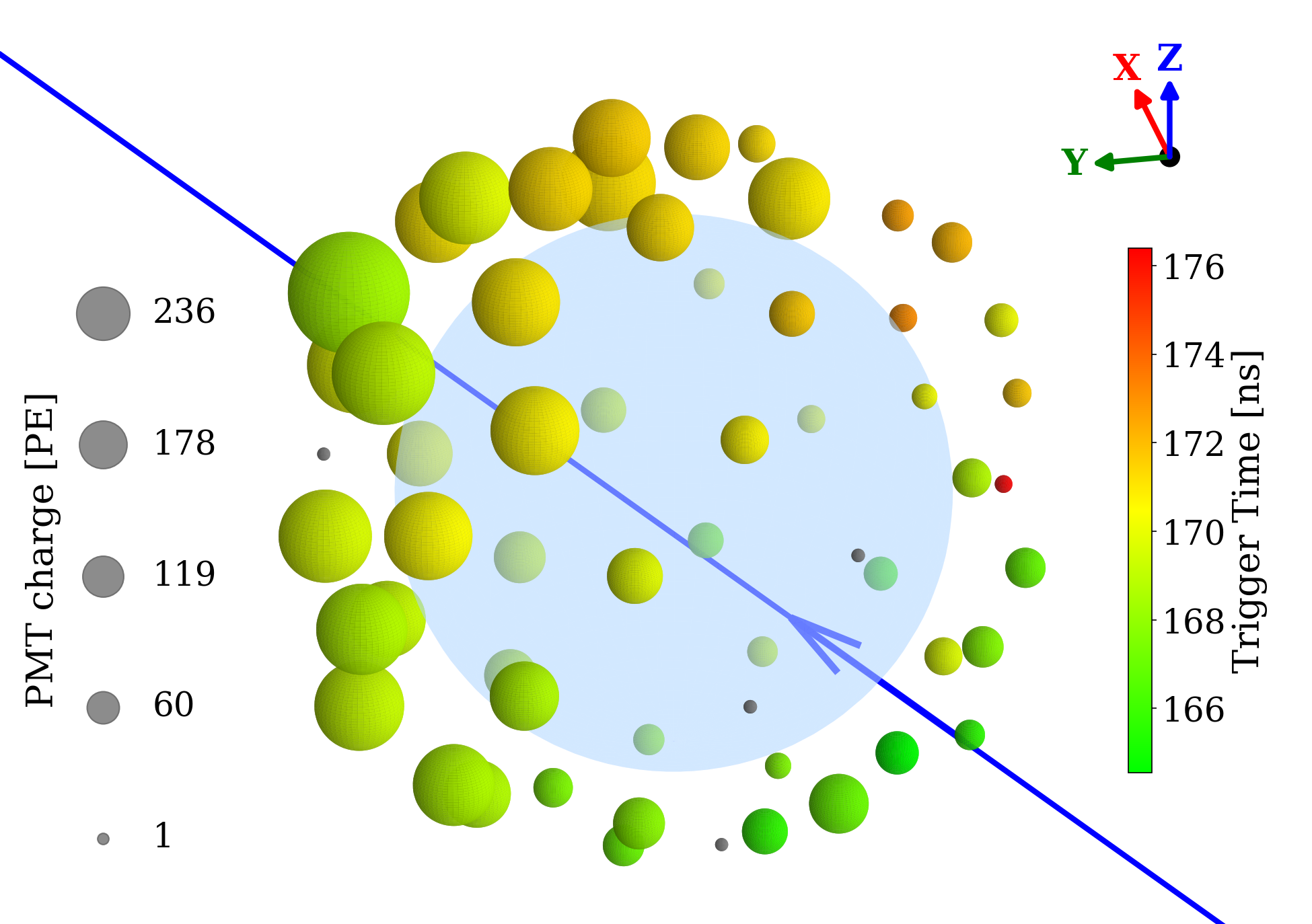}
\caption{Event display of the up-going muon candidate (run 48264, trigger 1337001) reconstructed by the template method. The PMTs near the entry point are lit first, and the charge is largest near the exit point, as expected for a muon traversing the detector from bottom to top.
}
\label{fig:upgoing}
\end{figure}

Based on the measured up-going muon fluxes in Refs.~\cite{SuperK:1999UpgoingMuon, MACRO:1998UpgoingMuon, Baksan:1999UpgoingMuon, IceCube:2011AtmMuNu, ANTARES:2013AtmMuNu} and the cosmic-ray muon flux at CJPL~\cite{Zhang_2024_Neutron_CJPLI}, the prior ratio of neutrino-induced up-going muons to cosmic-ray muons is taken to be about $1/300$, assuming the same detector acceptance and detection efficiency for the two event classes; for the 71 muons selected in the two water phases this corresponds to an expected number of about $0.24$ up-going muons.
One up-going muon candidate was identified during the Water-II data taking. As noted above, this muon event is not included among the 45 Water-II muon events used for the flux measurement and the direction reconstruction. 
It has 56 triggered PMTs, with flasher-rejection variables $r_\mathrm{max}=0.08$ and $\mathrm{mrt}=6.53$~ns. The template-based reconstruction yields $\theta=125.4^\circ$, $\phi=27.0^\circ$, as shown in FIG.~\ref{fig:upgoing}. 

The up-going interpretation of this event is established through a series of independent checks. 
\begin{itemize}
    \item First, the event passes the complete muon selection chain, including the good-run selection, the flasher rejection, the triggered-PMT-multiplicity,  and the number of PE requirements; its charge and timing patterns are consistent with a particle rather than a PMT flasher or a dark-noise fluctuation.
    \item Secondly, the observed topological pattern of Cherenkov light is consistent with that expected for upward‑going events; a characteristic feature of muon events traversing the detector is that PMTs near the incident point are triggered first, while the region with the largest charge is located near the exit point.
    \item Third, the template-based reconstruction returns a best-fit zenith angle above $90^\circ$, and all best-matching templates favor an upward trajectory. With the 68.3\% containment angle of about $6^\circ$ obtained in Sec.~\ref{sec:Rec}, the reconstructed zenith angle of $\theta = 125.4^\circ$ lies $35.4^\circ$ away from the horizontal, corresponding to $35.4^\circ/6^\circ \simeq 5.9\sigma$ if the containment angle is identified with the 1$\sigma$-equivalent angular error; this independently disfavors a downward-going interpretation. 
    \item Finally, the possibility that this candidate is a downward-going cosmic-ray muon mis-reconstructed as up-going is evaluated from the Monte Carlo sample. Requiring the same full-PMT-trigger condition as this candidate (56 triggered PMTs) and the same reconstructed direction region ($\cos\theta < -0.5$, i.e.\ $\theta > 120^\circ$), the mis-reconstructed-downward-muon hypothesis corresponds to a $p$-value of $1\times10^{-4}$, i.e.\ it is disfavoured at the 99.99\% confidence level ($3.7\sigma$).
\end{itemize}

At the 2400~m rock overburden of CJPL-I, the surviving atmospheric muons are essentially all downward-going. An up-going muon therefore cannot be produced by cosmic rays penetrating the overburden; instead, it is a characteristic signature of a neutrino-induced muon, as observed in water Cherenkov and underground experiments~\cite{Aharmim_2009_SNO_Muon, SuperK:1999UpgoingMuon, MACRO:1998UpgoingMuon, Baksan:1999UpgoingMuon, IceCube:2011AtmMuNu, ANTARES:2013AtmMuNu}, in which a muon neutrino (or antineutrino) interacts in the rock or water beneath the detector and produces a muon that propagates upward through the detector.

The identification of an up-going muon demonstrates, for the first time at CJPL, the capability of the JNE-1ton water Cherenkov detector to detect neutrino-induced events, opening a new window for neutrino-induced event studies at CJPL. 




\section{\label{sec:Sum} Summary}
This paper presents the first measurement of the cosmic-ray muon flux using a water Cherenkov detector at the CJPL-I. 
The upgraded JNE-1ton, equipped with 60 MCP-PMTs and operated for 75.25 and 69.88 live days in the two water-phase data-taking periods (Water-I and Water-II), has demonstrated reliable performance in muon detection and direction reconstruction.

The muon flux measured with the water mode JNE-1ton is
\begin{equation}
 \phi_{\text{I+II}} = (3.55 \pm 0.43_{\mathrm{stat}}\pm 0.28_{\mathrm{syst}}) \times 10^{-10}~\mathrm{cm}^{-2}\mathrm{s}^{-1} 
\end{equation}
which is in good agreement with the previous JNE-1ton result from 2017 to 2023~\cite{Zhang_2024_Neutron_CJPLI}. 
Compared with the previous LS JNE-1ton, the Water-II improves the muon detection efficiency by about 59\%, mainly owing to the larger effective detection volume inside the black shield. 

In the Water-I phase, the dominant systematic uncertainty arises from the effects of the water absorption length and the PMT photon detection efficiency on the total number of hit PMTs in muon events, leading to a muon flux uncertainty as large as 21.7\%. In the Water-II phase, thanks to the black shield, the muon selection is based on energy rather than on the number of hit PMTs, ultimately reducing the total systematic uncertainty to 7.5\%.

The optimized template‑based reconstruction method incorporates both charge and timing information, and employs a marginal‑likelihood approach to handle sparse hit patterns in the detector. This method improves the muon directional resolution to approximately 6$^{\circ}$. 
For events with 50 or more triggered PMTs, the achieved precision is roughly 3.4 times better than that of the original method (i.e., the angular error is reduced to $\simeq$30\% of its previous value).

Furthermore, one up-going muon candidate was identified in the Water-II data; its charge and timing topology and reconstructed direction are consistent with a neutrino-induced up-going muon. This is the first such event observed at CJPL, demonstrating the capability of the water Cherenkov detector for neutrino-induced event studies.

This work marks the first time that CJPL has employed a cost‑effective water detector for muon flux measurement, providing an independent cross-check of the existing liquid scintillator results and an important validation of the water Cherenkov detection technique.


\begin{acknowledgments}
This work was supported in part by the National Natural Science Foundation of China (NSFC) under Grant 12127808 and 12505130, the Ministry of Science and Technology of China (No. 2022YFA1604704), and the China Postdoctoral Science Foundation (Certificate Number: 2024M751611).
We would like to thank CJPL and its staff for hosting and supporting the JNE project. CJPL is jointly operated by Tsinghua University and Yalong River Hydropower Development Company.
\end{acknowledgments}



\bibliography{refs}

\end{document}